\documentclass[useAMS,usenatbib,fleqn]{mn2e}

\pdfoutput=1

\usepackage{graphicx,amsmath,multirow,hyperref,booktabs,units,amssymb}

\usepackage{times}

\newcommand{\edit}[1]{{#1}}

\title[Kinematic active region formation]{Kinematic active region formation in a three-dimensional solar dynamo model}
\author[A. R. Yeates and A. Mu\~{n}oz-Jaramillo]{A. R. Yeates$^{1}$\thanks{E-mail:
anthony.yeates@durham.ac.uk (ARY); amunoz@cfa.harvard.edu (AM-J)} and A. Mu\~{n}oz-Jaramillo$^{2,3,4}$\\
$^{1}$Department of Mathematical Sciences, Durham University, South Road, Durham, DH1 3LE\\
$^{2}$Harvard-Smithsonian Center for Astrophysics, Cambridge, MA 02138, USA\\
$^{3}$University Corporation for Atmospheric Research, Boulder, CO 80307, USA\\
$^{4}$Department of Physics \& Astronomy, University of Utah, Salt Lake City, UT 84112, USA}

\begin{document}

\date{Accepted ???? December ??. Received ???? December ??; in original form ???? October ??}

\pagerange{\pageref{firstpage}--\pageref{lastpage}} \pubyear{2013}

\maketitle

\label{firstpage}

\begin{abstract}
We propose a phenomenological technique for modelling the emergence of active regions within a three-dimensional, kinematic dynamo framework. By imposing localised velocity perturbations, we create emergent flux-tubes out of toroidal magnetic field at the base of the convection zone, leading to the eruption of active regions at the solar surface.  The velocity perturbations are calibrated to reproduce observed active region properties (including the size and flux of active regions, and the distribution of tilt angle with latitude), resulting in a more consistent treatment of flux-tube emergence in kinematic dynamo models than artificial flux deposition. We demonstrate how this technique can be used to assimilate observations and drive a kinematic 3D model, and use it to study the characteristics of active region emergence and decay as a source of poloidal field.  We find that the poloidal components are strongest not at the solar surface, but in the middle convection zone, in contrast with the common assumption that the poloidal source is located near the solar surface.  We also find that, while most of the energy is contained in the lower convection zone, there is a good correlation between the evolution of the surface and interior magnetic fields.
\end{abstract}


\section{Introduction}

The emergence of bipolar magnetic regions (BMRs) on the solar photosphere, with varying frequency and location, is one of the main signatures of the solar cycle.  These regions are characterised by a highly complex magnetic field whose catastrophic relaxation leads to violent releases of energy, making them the main source behind changes in the interplanetary environment (commonly referred to as space weather; \citealt{Schwenn2006}).  Besides their important role for determining the evolution of the magnetic field in the heliosphere, the collective effect of BMR emergence and decay on the recreation of the large-scale poloidal field is believed to be a crucial link in the progression of the solar magnetic cycle  (a process also known as the Babcock-Leighton, BL, mechanism; \citealt{Babcock1961,Leighton1969}).  For this reason, understanding BMR dynamics is crucial for understanding the cycle itself.

BMRs are believed to result from emergence of flux-ropes originating at the bottom of the convection zone \citep{Fan2009}, where magnetic field is thought to be stored and amplified as part of the dynamo mechanism \citep{Charbonneau2010}.  However, due to the lack of direct magnetic field measurements inside the convection zone, the mechanisms behind the formation and emergence of these flux-tubes are not yet fully understood.  So far two complementary approaches have been used to improve our understanding of flux-tube dynamics.  The first approach uses the thin flux-tube approximation, where all physical quantities of the tube are assumed to be averages over the tube cross-section. This approach takes advantage of the fact that, in most of the convection zone, the length-scales involved in flux-tube evolution are much larger than the tube's cross section \citep{Spruit1981}.  This limits the scope of these simulations to points along the flux-tube, greatly reducing the computational cost.  Thin flux-tube simulations have been instrumental in developing the current picture of flux-tube emergence and identifying the factors that determine the properties of BMRs \citep{DSilva1993,Schussler1994,Caligari1998,Fan2000,Weber2011}.  The second approach is the numerical solution of the anelastic magneto-hydrodynamic (MHD) equations \citep{Fan2008,Nelson2011b,Jouve2013,Pinto2013,Fan2013}.  These are more detailed simulations in which the interaction between the flux-tube and turbulent convection is studied directly, while filtering out sound waves to make the simulation feasible.  Though highly computationally intensive, anelastic MHD simulations are able to capture better the dynamics of the upper convection zone, where the thin flux-tube approximation is no longer accurate.

Due to the limited scope of thin flux-tube models and the cost of anelastic MHD simulations, most of the modelling effort has focused on reproducing the general properties of BMRs (tilt, latitude of emergence, rise time, etc.) and the details of flux-tube emergence -- finishing the simulation once the flux-tube reaches near surface layers (0.96-0.98 solar radii).  However, the study of multiple flux-tube emergence in the context of solar cycle propagation, also including flux-tube decay, has not received the same amount of attention.  
Typically, solar cycle models use a kinematic formulation based on the ideal induction equation, where magnetic flux transport is governed by advection and turbulent diffusion processes (see review by \citealt{Charbonneau2010} and references therein).  Currently, the closest approximation used to model BMR emergence, in this type of models, is to artificially deposit bipolar magnetic structures near the surface based on the toroidal field in the bottom of the convection zone \citep{Durney1997,Nandy2001,Munoz-Jaramillo2010,Guerrero2012,Hazra2013}.  However, this approach bypasses the process of flux-tube emergence, resulting in a physical disconnection between the magnetic field of BMRs and the toroidal field belts in the lower convection zone. This makes flux conservation difficult to enforce and alters the process by which toroidal field belts repair after flux-tube emergence \citep{vanBallegooijen2007}.

When the emergence and decay of BMRs was first proposed as a mechanism for poloidal field generation \citep{Babcock1961,Leighton1969}, it was as part of a shallow dynamo which would take place exclusively near the surface.  However, after thin flux-tube simulations demonstrated the need for super-equipartition fields \cite[in order to explain the observed properties of BMRs; see review by][]{Fan2009}, the original BL idea was transformed from a shallow dynamo (in which the the transition between poloidal and toroidal fields is restricted to the surface), to a double-layer dynamo in which the toroidal field is amplified and stored in a stable layer beneath the convection zone, but in which the regeneration of the poloidal field remains restricted to the surface.

This physical separation of the toroidal and poloidal sources is very appealing from a theoretical point of view, because understanding the solar cycle becomes a matter of pinning down the flux-transport mechanisms that communicate between the two source layers. Nevertheless, the assumption of a poloidal source confined to the surface is mainly a consequence of the historical development of the BL idea and, to this date, has never been substantiated because this requires a three-dimensional simulation featuring realistic buoyant eruptions in sufficient numbers over cycle timescales.

In this paper we propose a new technique that takes advantage of the kinematic framework to model the full process of three-dimensional flux-tube emergence in the context of solar dynamo models.  This technique is designed to incorporate the key features of emerging flux-tubes, as determined by thin flux-tube and anelastic MHD simulations, and allows for a more consistent treatment of flux-tube emergence in kinematic dynamo models than artificial flux deposition.  Additionally, we use this technique to improve our understanding of the emergence and decay of BMRs as a source for creating poloidal field out of toroidal field (i.e. the BL mechanism); by taking advantage of a significant reduction in the amount of necessary assumptions made by other algorithms for BMR deposition (specially regarding BMR shape and extent into the convection zone).  Finally, we demonstrate how our technique can be used to assimilate BMR observations and drive a dynamo model with aims to better understand observed cycle properties and to seed future model-based predictions.

The layout of this paper is as follows. Our new emergence technique is described in Section \ref{sec:model}, while the background velocity and diffusion profiles used in our simulations are set out in Section \ref{sec:flows}. Simulations of individual flux tubes are discussed in Section \ref{sec:tubes}, and a full solar cycle simulation is presented and analysed in Section \ref{sec:cycle}. Conclusions are summarised in Section \ref{sec:conclusions}. Details of the numerical methods used for our three-dimensional  simulations are given in Appendix A.

\section{Flux-tube Emergence Method} \label{sec:model}

\begin{figure}
\centering
\includegraphics[width=60mm]{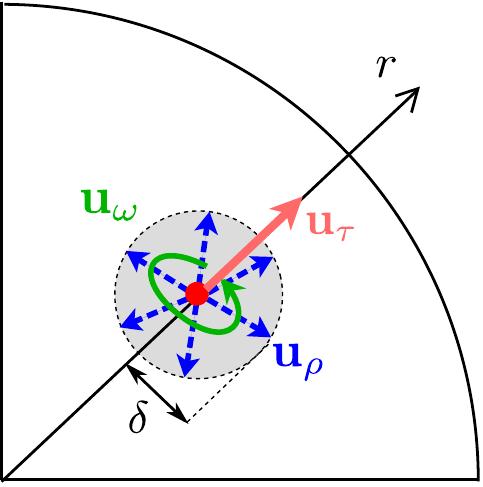}
\caption{Geometry of a velocity perturbation centred at $(\bar{r},\bar{\theta},\bar{\phi})$. \label{fig:geom}}
\end{figure}

Kinematic dynamo models are based on the ideal MHD induction equation
\begin{equation}
\frac{\partial{\bmath B}}{\partial t} = \nabla\times({\bmath v}\times{\bmath B}) - \nabla\times(\eta\nabla\times{\bmath B}),
\label{eqn:inductionB}
\end{equation}
where the magnetic field ${\bmath B}$ evolves according to the prescribed velocity ${\bmath v}$ and turbulent diffusivity $\eta$.
Typical velocity fields used in solar cycle simulations include the axisymmetric effects of differential rotation ${\bmath v}_\Omega$, meridional flow ${\bmath v}_M$ and turbulent pumping ${\bmath v}_{\rm P}$. Recent thin flux-tube simulations, embedded in a turbulent convective background taken from anelastic MHD simulations, have found that the process of flux-tube emergence is strongly determined by the tube's interaction with the convective flows \citep{Weber2011}. To approximate this in the kinematic framework, we add a time-dependent non-axisymmetric velocity perturbation ${\bmath u}$ that models the combined effect of buoyancy and turbulent advection on flux-tubes. This perturbation ${\bmath u}$ is localised in space and time, and includes three basic components (see Fig. \ref{fig:geom}):
\begin{enumerate}
\item An outward radial component ${\bmath u}_\tau$ responsible for transport of the tube from the base of the convection zone to the surface.
\item A vortical component ${\bmath u}_\omega$ that captures the net effect of helical turbulence on the rising tube, imparting a tilt to the resulting photospheric BMR.
\item A diverging component ${\bmath u}_\rho$ that expands the tube as it rises, in accordance with the decrease in surrounding density.
\end{enumerate}

Next, we give the detailed expressions used for the components of ${\bmath u}$, in spherical polar coordinates $(r,\theta,\phi)$. We denote the tube centre by $(\bar{r}(t),\bar{\theta}(t),\bar{\phi}(t))$, and write the Euclidean distance of a point $(r,\theta,\phi)$ from this centre as
\begin{equation}
\xi = \sqrt{r^2 + \bar{r}^2 - 2r\bar{r}\Big(\sin\theta\sin\bar{\theta}\cos(\phi-\bar{\phi}) + \cos\theta\cos\bar{\theta} \Big)}.
\end{equation}
The velocity perturbation centre is assumed to move radially at constant speed $u_0$, and also in longitude with the local differential rotation. We ignore the effect of advection by meridional circulation and turbulent pumping on the perturbation centre, as these are negligible on the emergence time-scale. The position at each timestep is determined from
\begin{equation}
\frac{d\bar{r}}{dt} =  u_0, \qquad \frac{d\bar{\theta}}{dt} = 0,\qquad \frac{d\bar{\phi}}{dt} = \Omega(\bar{r}(t),\bar{\theta}(t)).
\end{equation}
The speed $u_0$ is set to give 25 days travel time from $r=0.7R_\odot$ to $r=R_\odot$ (where $R_\odot = 6.96\times10^{10}$cm is the solar radius), and once the center of the perturbation reaches the photosphere, the perturbation is removed.

\subsection{Outward radial component}

The outward radial component of the velocity perturbation takes a three-dimensional Gaussian form centred at $(\bar{r},\bar{\theta},\bar{\phi})$, namely
\begin{equation}
{\bmath u}_\tau = u_0e^{-\xi^2/\delta^2}{\hat{\bmath e}}_r.
\end{equation}
The parameter $\delta$ represents the radius of the velocity field ${\bmath u}$, which controls the ultimate size of the photospheric BMR.

\subsection{Vortical component}

The ${\bmath u}_\omega$ component takes the form of an incompressible azimuthal velocity $\omega_1(r/2)e^{-\xi^2/\delta^2}$ around the radial vector, again centred at $(\bar{r},\bar{\theta},\bar{\phi})$. In spherical coordinates this has components
\begin{align}
u_{\omega\theta}&=-\frac{\omega_1r}{2}e^{-\xi^2/\delta^2}\sin\bar{\theta}\sin(\phi-\bar{\phi}),\\
u_{\omega\phi}&=\frac{\omega_1r}{2}e^{-\xi^2/\delta^2}\Big(\sin\theta\cos\bar{\theta} - \cos\theta\sin\bar{\theta}\cos(\phi-\bar{\phi}) \Big).
\end{align}
For the angular velocity of the helical motion, we set
\begin{equation}
\omega_1 = -\omega_0\cos\bar{\theta},
\end{equation}
where the cosine factor models the effect of the Coriolis force, giving in particular the opposite sign in each hemisphere, and the constant $\omega_0$ is calibrated to match the observed Joy's Law (see Section \ref{sec:params}).

\subsection{Diverging component}

\begin{figure}
\includegraphics[width=84mm]{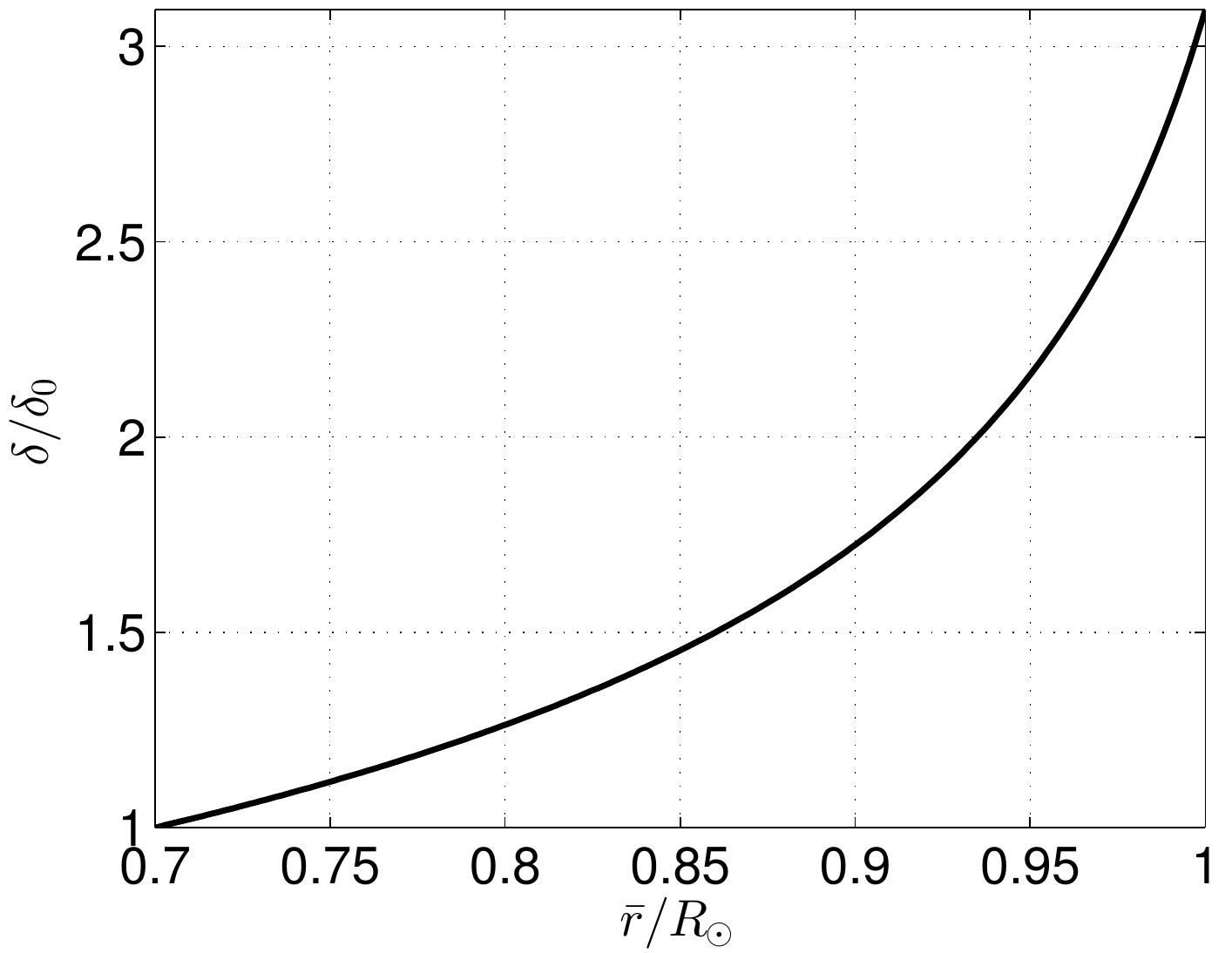}
\caption{Flux-tube radius $\delta$ as a function of tube centre radius $\bar{r}$. \label{fig:delta}}
\end{figure}

Expansion of the rising tube is achieved in two ways: firstly the tube radius $\delta$ is increased as $\bar{r}$ increases, and secondly a diverging flow ${\bmath u}_\rho$ is applied within the tube. We assume an adiabatic expansion of the tube so that $\rho/\rho_0=\delta^3/\delta_0^3$, where $\rho_0$, $\delta_0$ are initial values of the background plasma density and tube radius at $\bar{r}=\bar{r}_0$, and $\rho$, $\delta$ are values at some later time. Then assuming a density profile
\begin{equation}
\rho(r) = \left(\frac{R_\odot}{r} - 0.95 \right)^{3/2}
\end{equation}
we find
\begin{equation}
\delta(\bar{r})=\delta_0\sqrt{\frac{R_\odot/\bar{r}_0 - 0.95}{R_\odot/\bar{r} - 0.95}}.
\label{eqn:delta}
\end{equation}
As shown in Fig. \ref{fig:delta}, this leads to an approximately three-fold increase in tube radius between the base of the convection zone and the surface.

To define ${\bmath u}_\rho$, we impose a diverging flow of the form
\begin{equation}
{\bmath u}_\rho = u_{\rho0}\left(\frac{\xi}{R_\odot}\right)\frac{1}{2}\left(1-{\rm erf}\,\left(\frac{\xi-\delta}{0.2\delta}\right) \right){\hat{\bmath e}}_\xi,
\end{equation}
where ${\hat{\bmath e}}_\xi$ is a radial unit vector centred at $(\bar{r},\bar{\theta},\bar{\phi})$. The magnitude of this expansion, $u_{\rho0}$ must be chosen to match the expansion rate of the tube radius $\delta$ in Equation \eqref{eqn:delta}. To motivate our choice, consider the perpendicular expansion of a uniform vertical field $B_z$. The induction equation requires that the expansion velocity satisfies
\begin{equation}
v_R = -\frac{R}{2B_z}\frac{\partial B_z}{\partial t},
\label{eqn:vdiv}
\end{equation}
where $R$ is the radial vector in the plane perpendicular to $B_z$. If we require $B_z$ to remain constant in space, decreasing in time as the tube expands, then flux conservation in the tube requires $B_z\delta^2=B_{z0}\delta_0^2$, or
\begin{equation}
B_z = B_{z0}\left(\frac{R_\odot/\bar{r} - 0.95}{R_\odot/\bar{r}_0 - 0.95}\right).
\end{equation}
Differentiating and substituting into equation \eqref{eqn:vdiv} shows that the required velocity is
\begin{equation}
v_R = \frac{(R/R_\odot)u_0}{2(\bar{r}/R_\odot)^2(R_\odot/\bar{r} - 0.95)}.
\end{equation}
Setting $R$ to $\xi$ gives our chosen magnitude
\begin{equation}
u_{\rho0}= \frac{u_0}{2(\bar{r}/R_\odot)^2(R_\odot/\bar{r} - 0.95)}.
\end{equation}

\section{Background Model Ingredients} \label{sec:flows}

\begin{figure*}
\includegraphics[width=140mm]{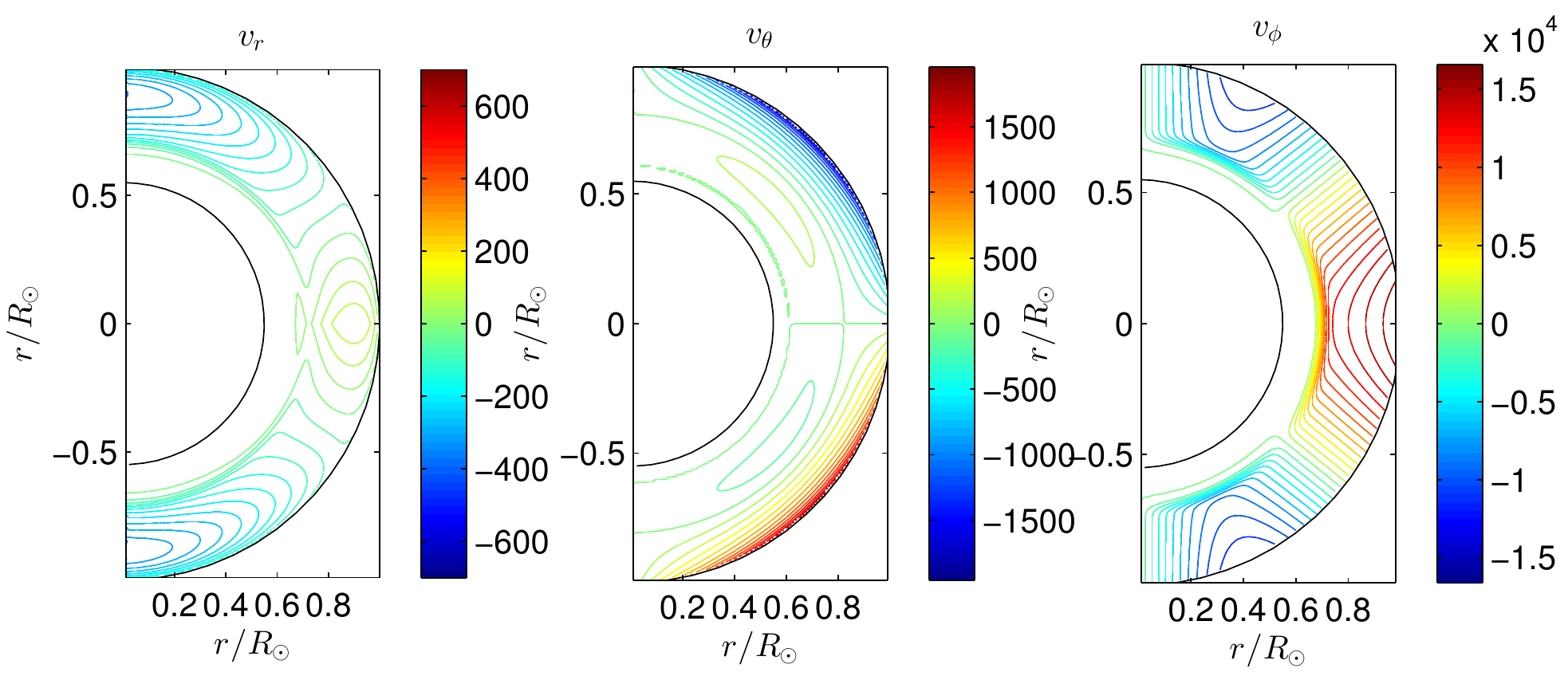}
\caption{Background velocity profiles, including differential rotation, meridional flow and turbulent pumping. The colour axis of each panel is in units of ${\rm cm}\,{\rm s}^{-1}$.\label{fig:vBG}}
\end{figure*}

Our model includes representative background velocity and diffusion profiles used in other solar cycle simulations. The background velocity includes contributions from differential rotation, meridional flow and turbulent pumping. The combined velocity components from all three contributions are illustrated in Fig. \ref{fig:vBG}.

\subsection{Differential rotation}

This takes the analytical form of \citet{charbonneau1999} with angular velocity
\begin{align}
\Omega(r,\theta)&=\Omega_C + \frac{1}{2}\left(1 + {\rm erf}\left( \frac{r-R_0}{\Delta_0}\right)\right)\cdot\nonumber\\
&\cdot \Big(\Omega_E-\Omega_C+(\Omega_P-\Omega_E)\left(C\cos^2\theta + (1-C)\cos^4\theta \right) \Big),
\end{align}
where $\Omega_C=2.71434\times 10^{-6}\,{\rm s}^{-1}$ is the angular velocity of the rigidly-rotating core, $\Omega_P=2.07345\times 10^{-6}\,{\rm s}^{-1}$ is that of the poles, and $\Omega_E=2.9531\times 10^{-6}\,{\rm s}^{-1}$ is that of the equator. The surface rotation profile is specified by $C=0.483$, and the tachocline lies at depth $R_0=0.7R_\odot$ with thickness $\Delta_0=0.025R_\odot$.

\subsection{Meridional flow}

We use a single-cell meridional flow in each hemisphere that is poleward at the surface and has an equatorward branch that penetrates beneath the tachocline \citep{Nandy2002a}. We set
\begin{equation}
{\bmath v}_M=\frac{1}{\rho(r)}\nabla\times\big(\Psi(r,\theta)\hat{\bmath e}_\phi\big),
\end{equation}
where the density profile is $\rho(r)=(R_\odot/r-0.95)^{3/2}$ and the streamfunction $\Psi(r,\theta)$ takes the form
\begin{equation}
\Psi(r,\theta) r\sin\theta=\frac{-v_0}{7.633}F(r)G(\theta),
\end{equation}
with
\begin{align}
\begin{split}
F(r) &= (r-R_p)\sin\left(\pi\frac{r-R_p}{R_\odot-R_p}\right)\exp\left[-\left(\frac{r-R_1}{\Gamma} \right)^2 \right],\\
G(\theta) &= \big(1-\exp(-1.5\theta^2) \big)\big(1 - \exp[1.8(\theta-\pi/2)]\big).
\end{split}
\end{align}
The constants are $R_p=0.62R_\odot$, $R_1=0.1125R_\odot$, $\Gamma=3.47\times 10^8\,{\rm m}$.

\subsection{Turbulent pumping}
The effect of turbulent pumping in kinematic dynamo simulations has been discussed by \citet{Guerrero2008} and \citet{Karak2012g}. We include pumping in the radial and latitudinal directions using the functional forms suggested by \citet{Kapyla2006b}. Following those authors, we neglect longitudinal pumping in comparison with differential rotation. The pumping contribution has the form of a velocity
\begin{align}
v_{{\rm P}\theta}&=\gamma_0\left({\rm tanh}\left(\frac{r-R_2}{\Delta_2} \right) - {\rm tanh}\left(\frac{r-R_3}{\Delta_3} \right) \right)\sin^4\theta\cos\theta,\\
v_{{\rm P}r}&=-\frac{\gamma_0}{4}\left({\rm tanh}\left(\frac{r-R_2}{\Delta_2} \right) - {\rm tanh}\left(\frac{r-R_4}{\Delta_4} \right)\right)\cdot\nonumber\\
&\cdot\left(1 + \exp\left(\frac{(r-R_2)^2}{\Delta_5^2}\right)|\cos\theta| \right)
\end{align}
where $R_2=0.71R_\odot$, $\Delta_2=0.015R_\odot$, $R_3=0.875R_\odot$, $\Delta_3=0.075R_\odot$, $R_4=0.975R_\odot$, $\Delta_4=0.1R_\odot$, $\Delta_5=0.25R_\odot$. The main features are (i) downward radial pumping at all latitudes, weaker nearer the equator, and (ii) equatorward latitudinal pumping, strongest near the base of the convection zone around $15\degr$ latitude. Both components vanish in the overshoot region. We take $\gamma_0=100\,{\rm cm}\,{\rm s}^{-1}$.

\subsection{Turbulent diffusivity}

We adopt an axisymmetric double-step radial profile that was chosen by \citet{Munoz-Jaramillo2011} to approximate the effect of magnetic quenching on turbulent diffusivity, namely
\begin{align}
\begin{split}
\eta(r)&=\eta_C + \frac{\eta_0-\eta_C}{2}\left(1 + {\rm erf}\left(\frac{r-R_5}{\Delta_6} \right) \right)  + \\
& + \frac{\eta_S - \eta_0 - \eta_C}{2}\left(1 + {\rm erf}\left(\frac{r-R_6}{\Delta_7}\right) \right).
\end{split}
\end{align}
Here $\eta_C=10^8\,{\rm cm}^2\,{\rm s}^{-1}$ is the core diffusivity, $\eta_S=6\times 10^{12}\,{\rm cm}^2\,{\rm s}^{-1}$ the diffusivity at $r=R_\odot$, and $\eta_0=1.6\times10^{11}\,{\rm cm}^2\,{\rm s}^{-1}$ the diffusivity in the convection zone. The step locations and thicknesses are $R_5=0.71R_\odot$, $\Delta_6=0.03R_\odot$, $R_6=0.95R_\odot$, $\Delta_7=0.025R_\odot$.

\begin{figure*}
\includegraphics[width=140mm]{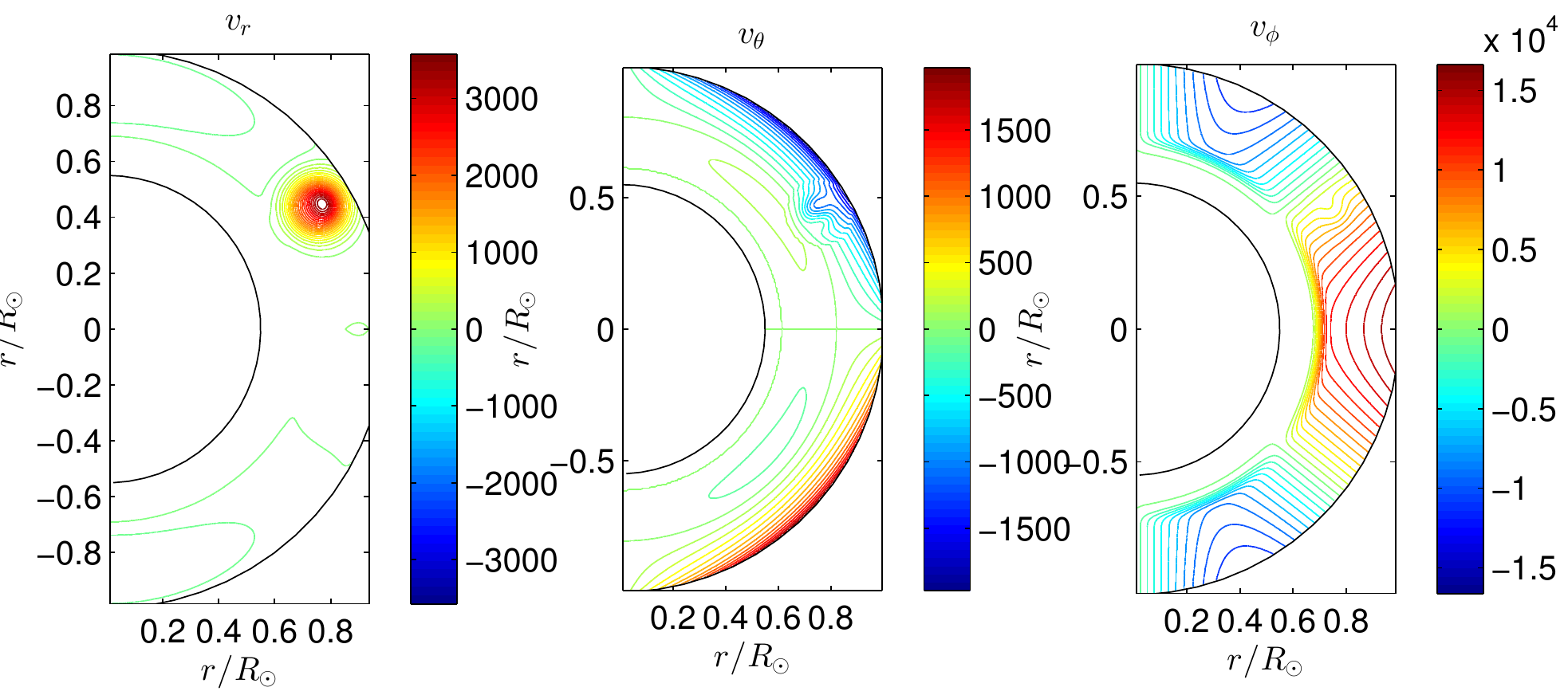}
\caption{Velocity perturbation associated with the flux-tube emergence at 30$\degr$N.  The meridional cuts are taken at $\phi=2\pi/3$ during day 15 of emergence. The colour axis of each panel is in units of ${\rm cm}\,{\rm s}^{-1}$. The velocity perturbation is embedded in the background velocity field shown in Fig. \ref{fig:vBG}.\label{fig:tubevel}}.
\end{figure*}

\begin{figure*}
\includegraphics[width=168mm]{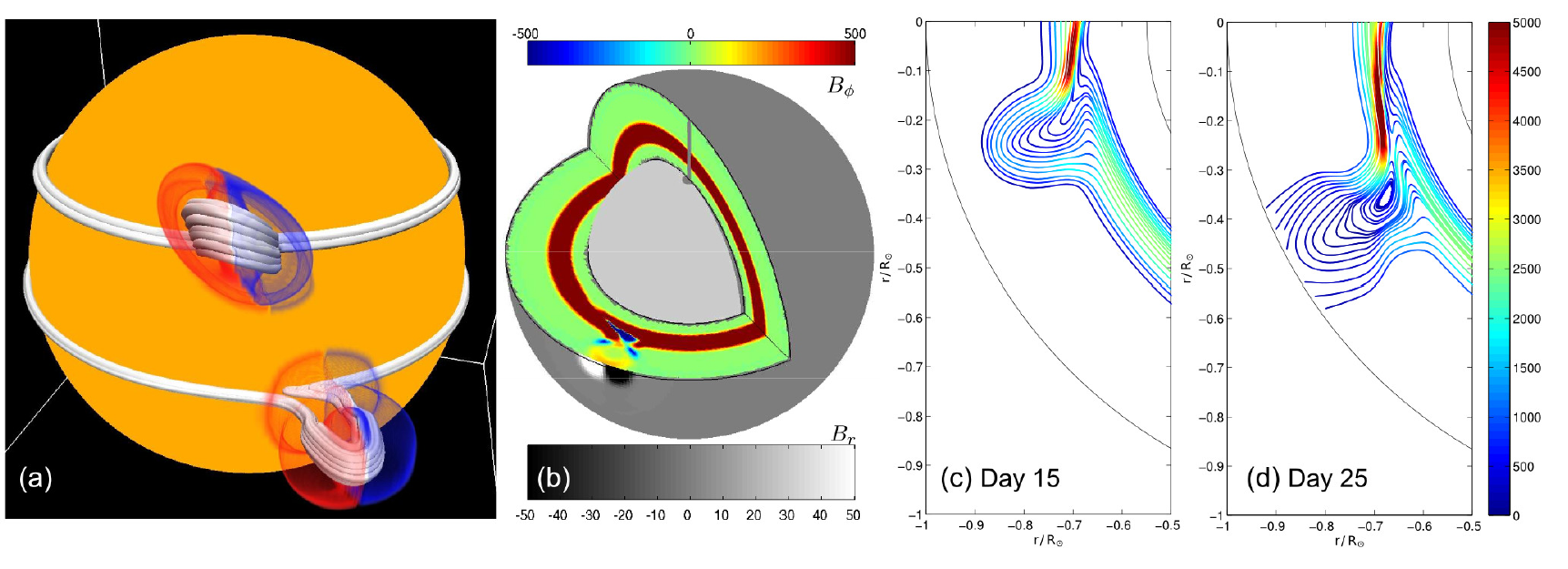}
\caption{Visualization of two isolated flux-tube emergences at $0\degr$ latitude and $30\degr$N on day 25 of emergence (a), and detail of the equatorial tube (b, c, d). In (a), coloured surfaces enclose regions of positive (red) and negative (blue) $B_r$. In (b), contours on the cuts show $B_\phi$, while those on the spherical surface show $B_r$ (also on day 25). Panels (c) and (d) show magnetic field-lines in the equatorial plane at on days 15 and 25, coloured by $|{\bmath B}|$. All colour axes are in units of Gauss. The simulation shown here used grid resolution $\Delta_\phi=2\pi/384$, $\Delta_r=0.45R_\odot/48$.
\label{fig:3d}}
\end{figure*}

\section{Simulation of Isolated Flux-Tubes} \label{sec:tubes}

We begin by simulating the emergence of isolated flux tubes, using the method introduced in Section \ref{sec:model}. In our simulations, we solve the ideal MHD induction equation (Eq.~\ref{eqn:inductionB}), expressed in terms of the vector potential, using a finite difference scheme (see Appendix A for details).

\subsection{Boundary and initial conditions} \label{sec:bcs}

For the simulations of isolated tubes, our initial conditions consist of a purely toroidal field layer in the tachocline, of the form
\begin{align}
{\bmath B}=\frac{B_0}{2}\left({\rm erf}\left(\frac{r-R_7}{\Delta_8}\right) - {\rm erf}\left(\frac{r-R_8}{\Delta_8}\right) \right){\hat{\bmath e}}_\phi,
\label{eqn:belt}
\end{align}
where $R_7=0.66R_\odot$, $R_8=0.74R_\odot$, $\Delta_8=0.018R_\odot$, and $B_0=2.5\times 10^3\,{\rm G}$.
For our lower boundary condition we assume a perfectly-conducting core, located at $R_{\rm min}=0.55R_\odot$, setting $\partial(rB_\theta)/\partial r=\partial(rB_\phi)/\partial r=0$.  For our upper boundary condition we assume a perfectly radial magnetic field ($B_\theta=B_\phi=0$), a condition found to be necessary for stress balance between subsurface and coronal magnetic fields \citep{vanBallegooijen2007}.

\subsection{Formation of bipolar magnetic regions}

\begin{figure*}
\includegraphics[width=168mm]{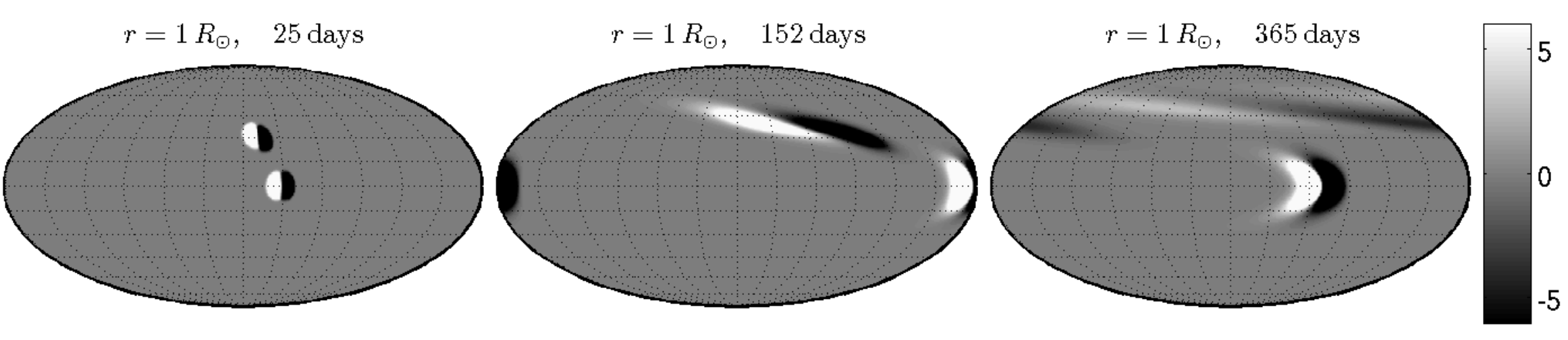}
\caption{Post-emergence evolution of $B_r$ on the photosphere for the two BMRs in Fig. \ref{fig:3d} (saturated at $\pm 6\,{\rm G}$). The full sphere is shown in Mollweide equal-area projection.
 \label{fig:brlong}}
\end{figure*}

\begin{figure*}
\includegraphics[width=168mm]{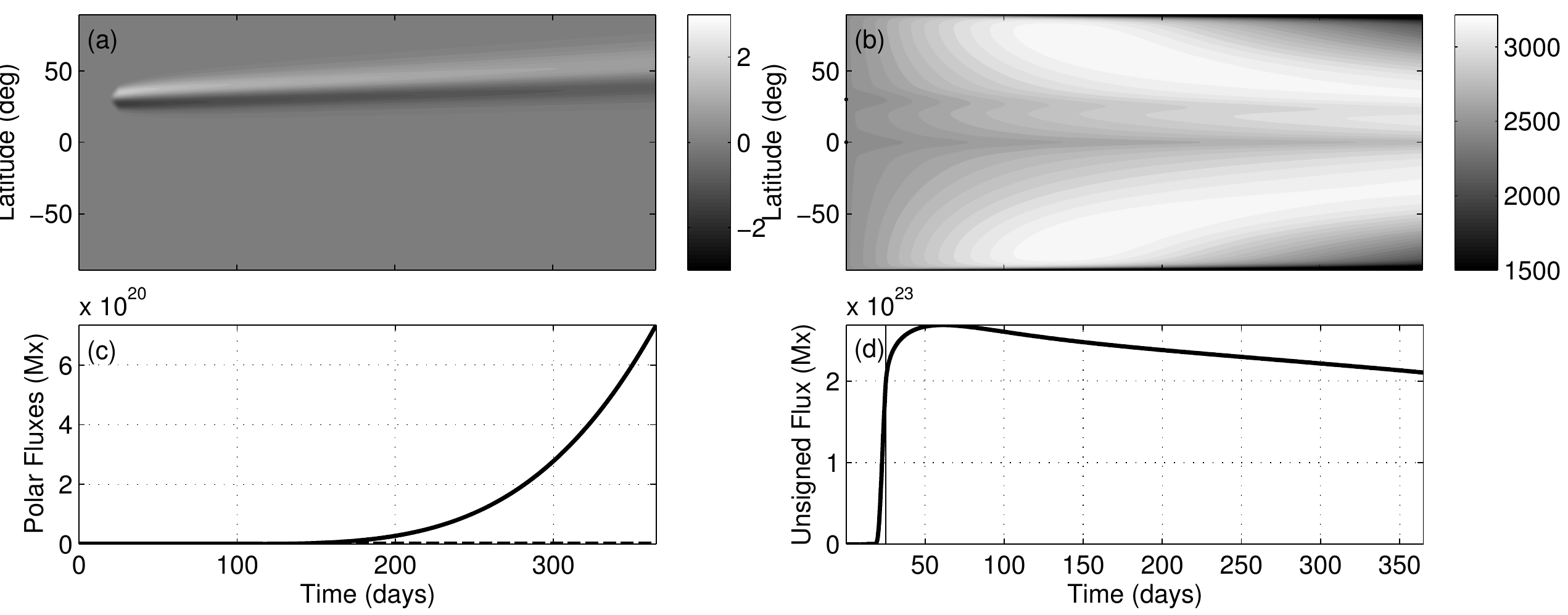}
\caption{Butterfly diagrams of longitude-averaged $B_r(R_\odot)$ (a) and $B_\phi(0.7R_\odot)$ (b), for the simulation with two BMRs, as shown in Fig. \ref{fig:3d}.
Panel (c) shows (signed) polar surface flux above $70\degr$ latitude, and panel (d) the total unsigned photospheric flux over all latitudes. \label{fig:bflylong}}
\end{figure*}

To illustrate the flux tube emergence process, we present a simulation with two simultaneously emerging flux tubes.  Both tubes share the same longitude, but one is located at the equator (with a co-latitude of $\bar{\theta}=\pi/2$) and the other at $30\degr$N  (with a co-latitude of $\bar{\theta}=\pi/3$).  Both are created by velocity perturbations with a width of $\delta_0=(5\pi/180)(0.7R_\odot)$, and both are initiated at $t=0$.  Figure \ref{fig:tubevel} shows the velocity perturbation associated with the emergence at $30\degr$N, during day 15 of emergence, embedded in the background velocity field. The velocity perturbation for both emergences is stopped after 25 days.

During the course of our simulation, it can be seen (Fig. \ref{fig:3d}) that the rotational shear of the emerging flux-tube (due to the radial gradient in differential rotation) leads to the relative movement of the flux-tube with respect to its roots.  This bends the leading leg of the flux-tube over the underlying sheath of toroidal field and stretches the trailing leg over the depleted area left over by the eruption.  This phenomenon is particularly noticeable near the equator where the radial shear is strongest, and becomes evident in Fig. \ref{fig:3d}-b as neighboring regions of opposite polarity toroidal field.  However, as a whole, this process generates no net toroidal flux, but rather leads to a topological change in the magnetic configuration near the eruption site.  Of particular interest is the fact that the net displacement of the flux-tube (and in particular the stretching of the trailing leg) leads to the pinching of the eruption site and the partial disconnection of the emergent flux-tube (see \ref{fig:3d}-d).  Although the eruption of flux-tubes in a kinematic framework is more a conceptual tool than a faithful reproduction of the conditions inside the solar convection zone, this phenomenon can also be observed in anelastic MHD simulations or rising flux-tubes (see fig. 5 of \citealt{Jouve2013}), and could be playing an important role in post-emergence connectivity.

Figure \ref{fig:brlong} shows the evolution of the surface magnetic field for the two simultaneous eruptions.  The BMR associated with the eruption at $30\degr$N is strongly sheared due to the action of differential rotation, whereas the equatorial eruption results in a BMR that conserves its integrity due to its lack of tilt and its location in a region of minimal latitudinal shear.  Due to its lack of tilt, the equatorial BMR shows no signature in a synoptic map of the longitudinally-averaged magnetic field (also known as magnetic butterfly diagram, see Fig. \ref{fig:bflylong}-a).  In contrast, the BMR associated with the eruption at $30\degr$N features prominently; as the signature associated with its leading and trailing polarities are transported towards the pole by turbulent diffusion and meridional flow.  The net effect is a build up of positive polar flux, evident in Fig. \ref{fig:bflylong}-c, and the contribution to the net dipolar moment that lies at the core of the BL mechanism.

In spite of not contributing to the axisymmetric evolution of the surface magnetic field, it is clear that both eruptions are contributing to the depletion of toroidal field ($B_\phi$) at the bottom of the convection zone (see Fig. \ref{fig:bflylong}-b).  This depletion persists for the duration of our simulation (one year) and drifts slightly equatorward due to the return meridional flow at this depth.  The evolution of total unsigned flux at the photosphere shows how most of this flux emerges at the surface within the 25 days set by our velocity perturbation (see Fig. \ref{fig:bflylong}-d)).  However, flux continues to emerge until about day 50 due to outward diffusion of field from beneath the surface.  In a full model with multiple emergence events, this process of flux removal and outward transport will contribute to the weakening of the old cycle's toroidal field so that it can be replaced with flux of the opposite sign for the new cycle.

\subsection{Calibration of observable properties} \label{sec:params}

In order to develop full solar cycle simulations, it is necessary to calibrate the parameters of our velocity perturbations so that the resulting BMRs at the photosphere match the properties of observed BMRs. For this purpose, we compare simulations of single flux tubes with measurements of 2211 BMRs obtained from the US National Solar Observatory/Kitt Peak (NSO/KP) synoptic radial-component magnetograms between 1996 and 2012 \citep[see][]{Yeates2007,Yeates2013b}. A semi-automated technique was used to identify the location, size, magnetic flux and tilt angle of the observed BMRs, upon their first appearance in the synoptic maps.

\begin{figure}
\includegraphics[width=82mm]{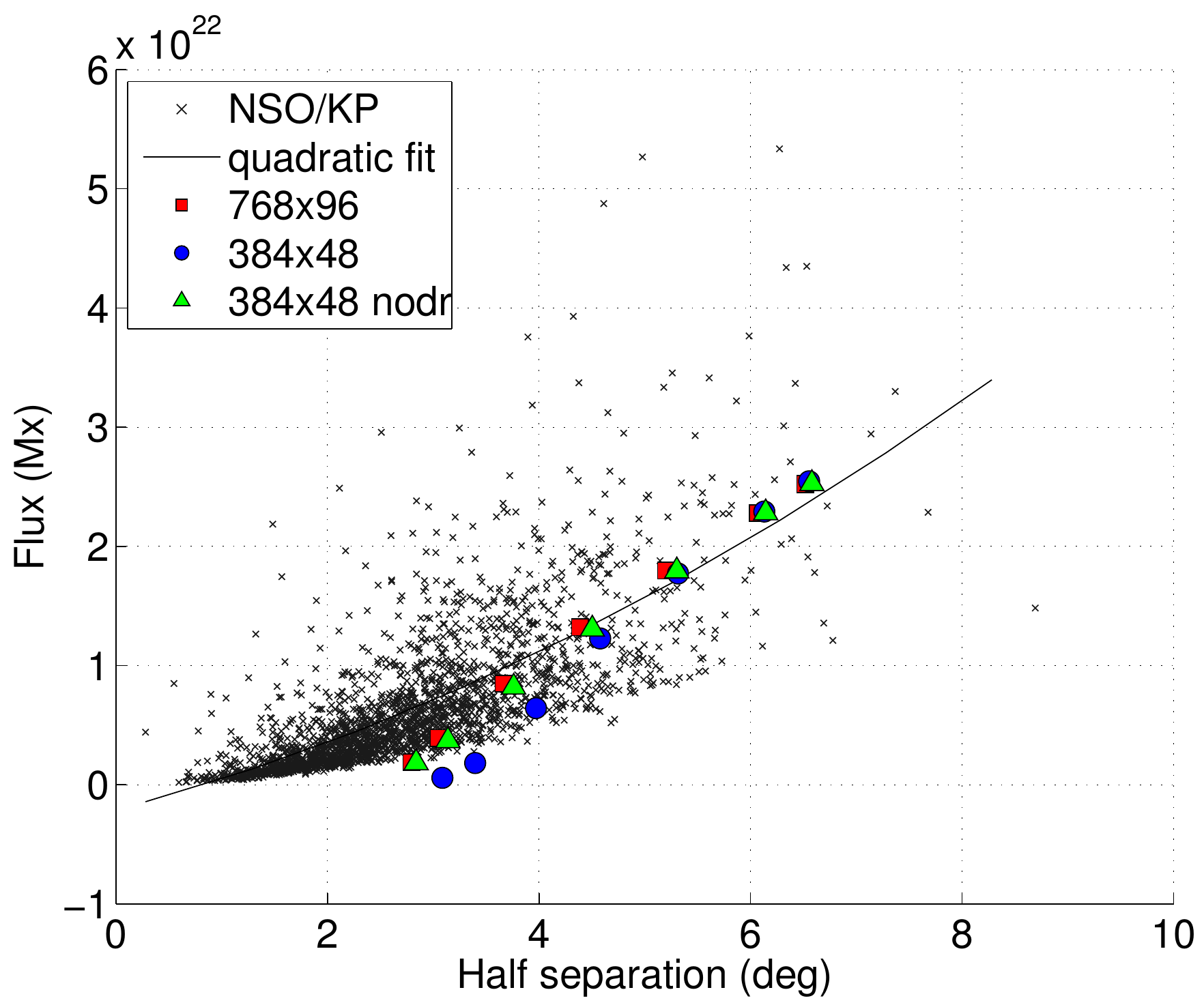}
\caption{Single-polarity flux against half-separation for BMRs. The underlying scatter plot shows observations from NSO/KP magnetograms, while larger symbols show different simulations. Simulations with red squares used resolution $\Delta_\phi=2\pi/768$, $\Delta_r=0.45R_\odot/96$, while simulations with blue circles or green triangles used $\Delta_\phi=2\pi/384$, $\Delta_r=0.45R_\odot/48$. In the simulations with green triangles, differential rotation was turned off. From left to right the simulations correspond to initial $\delta_0$ values of $\delta_0=\alpha(0.7R_\odot)$ with $\alpha=2.5\degr$, $3\degr$, $4\degr$, $5\degr$, $6\degr$,$7\degr$, $7.5\degr$. The solid line is a quadratic least-squares fit to the NSO/KP data.\label{fig:sepflux}}
\end{figure}

Firstly, we consider the size of the resulting photospheric BMRs. Figure \ref{fig:sepflux} shows the relation between size and flux of the BMRs, both for the observations (background scatter plot) and a series of different single tube simulations (large symbols). The flux shown in Fig. \ref{fig:sepflux} represents that of the leading polarity. The \emph{half-separation} is calculated as the half of the spherical angle between the centroid of the positive polarity ($\theta_p,\phi_p$ for $B_r>50\,{\rm G}$) and the centroid of the negative polarity ($\theta_n,\phi_n$ for $B_r<-50\,{\rm G}$). Defining the co-latitude of the BMR as $\theta_c=(\theta_p + \theta_n)/2$ and its longitude as $\phi_c=(\phi_p +\phi_n)/2$, then the half-separation $\rho$ satisfies
\begin{equation}
\cos\rho = \sin\theta_c\sin\theta_p\cos(\phi_c-\phi_p) + \cos\theta_c\cos\theta_p.
\end{equation}

The solid line in Fig. \ref{fig:sepflux} is a quadratic least-squares fit to the NSO/KP data, which gives $\Phi=(0.24\rho^2 + 2.34\rho - 2.10)\times10^{21}\,{\rm Mx}$. The flux in all simulations was scaled by setting $B_0=2500\,{\rm G}$. Then it can be seen that for $\delta_0\geq (5\pi/180)(0.7R_\odot)$, the simulated BMRs lie approximately on the observational best-fitting line. The BMRs created by smaller flux-ropes are seen to fall increasingly below this line. This is an effect of numerical resolution: the higher resolution runs (red squares) produce BMRs with smaller half-separation and greater flux, consistent with the error being caused by numerical diffusion in the advection scheme (Appendix A). To check this, we repeated the simulations with differential rotation turned off (green triangles), since this is our fastest flow. The consequent increase in BMR fluxes supports the interpretation that the error is caused by numerical diffusion, which would be proportional to the grid spacing and the velocity magnitude. This analysis shows that the smallest flux tubes adequately resolved by a grid with $\Delta_\phi=2\pi/384$, $\Delta_r=0.45R_\odot/48$ are about $\delta_0=(5\pi/180)(0.7R_\odot)$, which is adequate for a full-Cycle simulation using larger active regions (Section \ref{sec:cycle}).

Note that there is considerable scatter in the NSO/KP data in Fig. \ref{fig:sepflux}. The simulated BMRs all fall on a line because they were emerged from the same initial magnetic field. In a full simulation, the flux tubes will be created from differing initial field, which will lead to scatter.

\begin{figure}
\includegraphics[width=84mm]{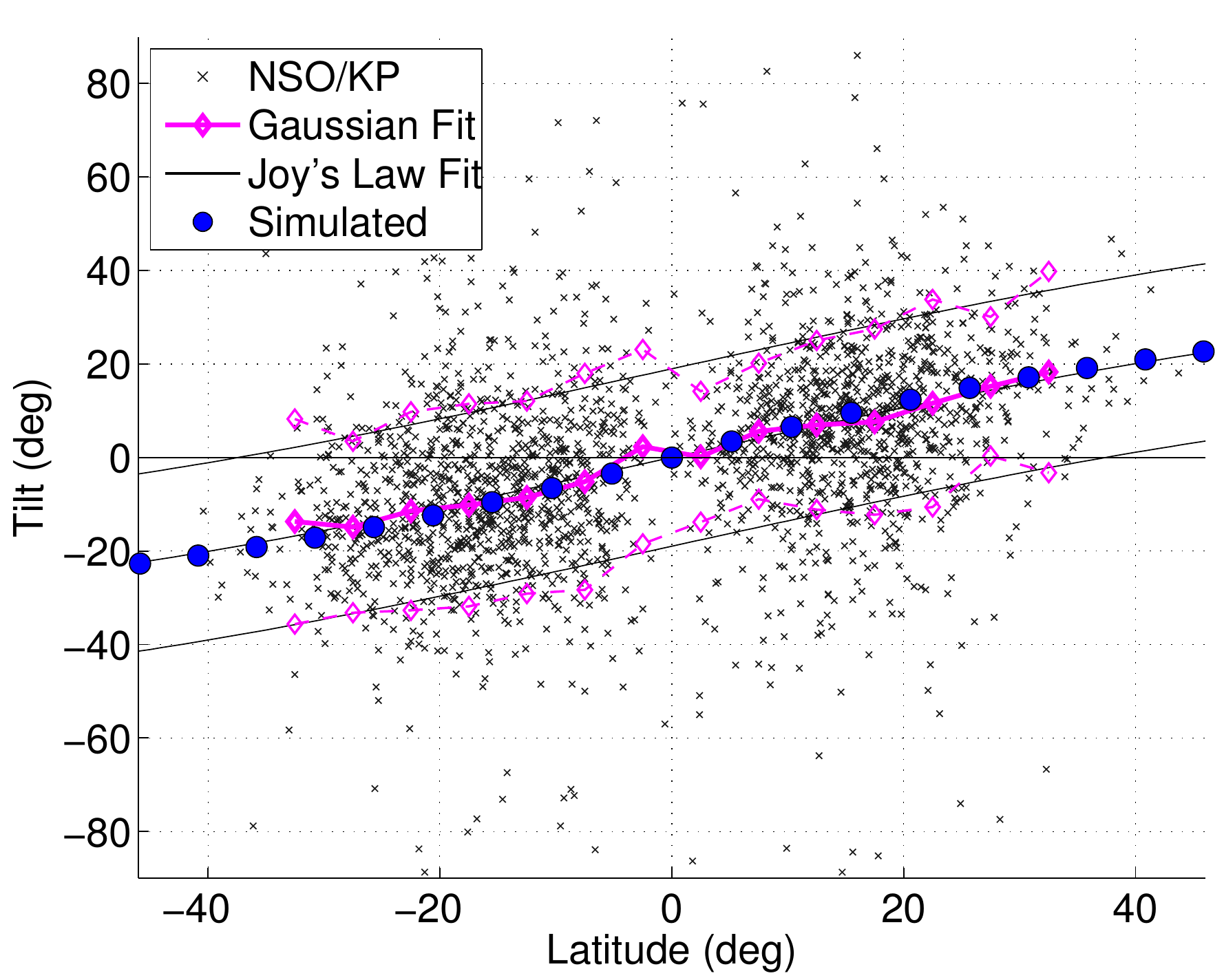}
\caption{Tilt angle against latitude for the BMRs. The underlying scatter plot shows observations from NSO/KP magnetograms (see text), while blue circles show the simulations at resolution $\Delta_\phi=2\pi/384$, $\Delta_r=0.45R_\odot/48$. The mean and standard deviation of the observed data in $5\degr$ latitude bins are shown in magenta. Thin black lines show analytical fits.  \label{fig:tilt}}
\end{figure}

Next, we turn to a property of BMRs that is critical for the BL mechanism: the tilt angle $\tau$ of their opposite photospheric polarities with respect to the equator, which we define as
\begin{equation}
\tan\tau = \frac{\theta_c - \theta_n}{\sin\theta_c(\phi_n-\phi_c)}.
\end{equation}
As described in Section \ref{sec:model}, the simulated BMRs acquire their tilt through the vortical component ${\bf u}_\omega$ of the flux tube velocity. Figure \ref{fig:tilt} shows the distribution of $\tau$ against latitude for the NSO/KP data, along with a series of simulated BMRs at different latitudes (blue circles). The magenta lines/symbols show the mean and standard deviation of fitted normal distributions to the NSO/KP data within $5\degr$ latitude bins (leaving out the outermost bins because they are too sparsely populated).  Although there is considerable scatter, the bin means show a clear latitudinal trend, which is well-established and known as Joy's Law \citep{Hale1919e}. A least-squares fit of the form $a\sin\lambda$ to these means gives $\langle\tau\rangle=0.55\sin\lambda$ (shown by the middle solid line), which is comparable to previous studies \citep{Wang1989a,Stenflo2012}.

The blue circles in Fig. \ref{fig:tilt} show the resulting tilt angles of photospheric BMRs in a sequence of simulations with flux tubes initiated at different latitudes. All of the flux tubes were given size $\delta_0=(5\pi/180)(0.7R_\odot)$. By repeating the simulations with varying $\omega_0$, the best-fitting value (illustrated here) of $\omega_0=0.08\times 10^{-5}\,{\rm s}^{-1}$ was selected to match the observed $\langle\tau\rangle$ curve. Using the same $\omega_0$ value for all flux tubes will obviously not reproduce the scatter evident in the observed tilt angles, but this could readily be incorporated in future simulations by statistically varying $\omega_0$ between tubes. To estimate the spread of the observed pattern, we note that the standard deviation is approximately equal in all latitude bins, with an average of $19.0\degr$. The outer solid lines in Fig. \ref{fig:tilt} show  $\langle\tau\rangle\pm 19\degr$.

\section{Simulation of a Full Solar Cycle} \label{sec:cycle}

\begin{figure*}
\includegraphics[width=168
mm]{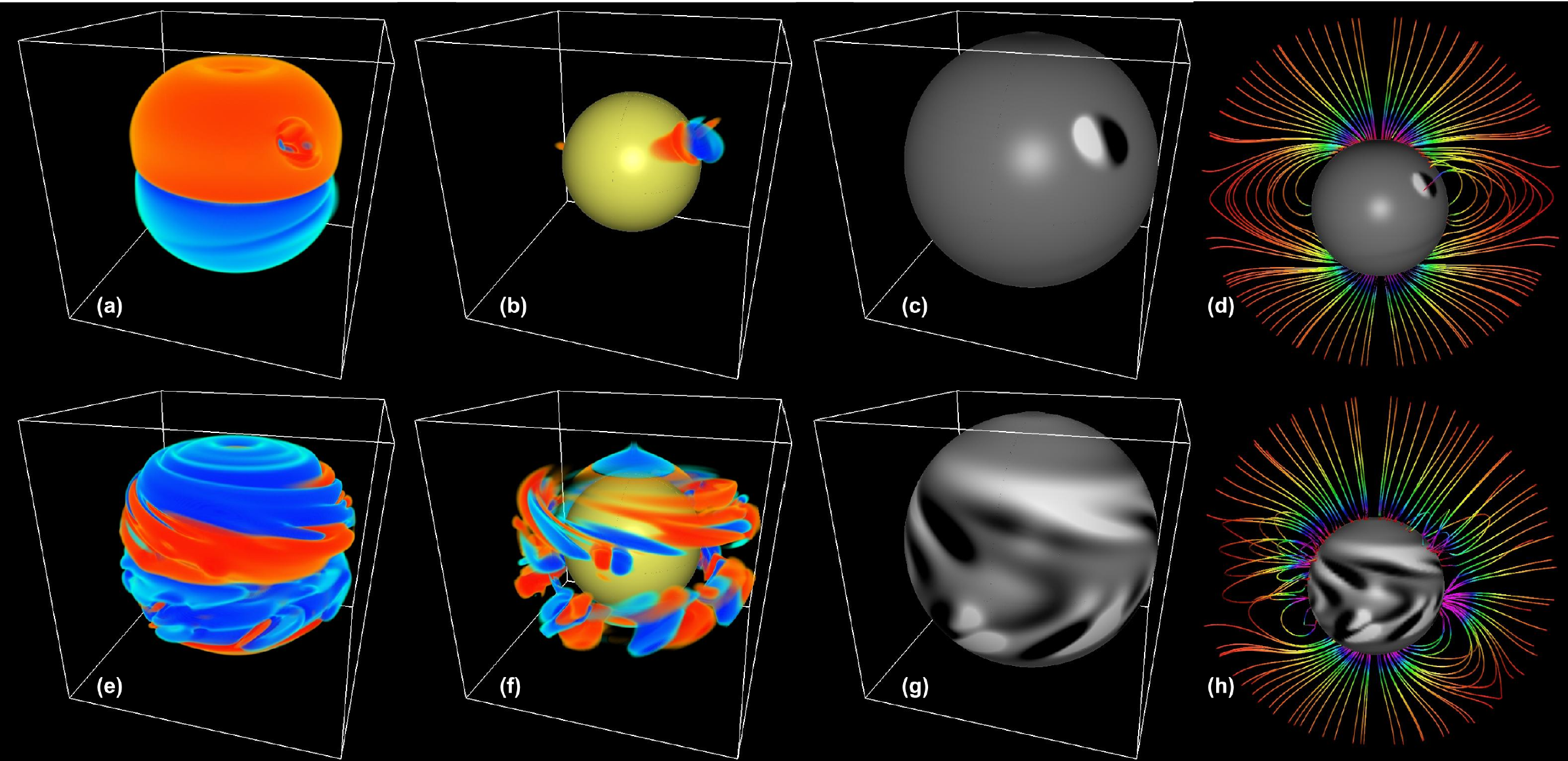}
\caption{Visualizations of ${\bmath B}$ in the full solar-cycle simulation: near cycle minimum (1 yr, panels a-d) and at cycle maximum (5.4 yr, panels e-h). Panels (a) and (e) show $B_\phi$ between $50-125\,{\rm G}$ (red positive, blue negative). Panels (b) and (f) show $\sqrt{B_r^2 + B_\theta^2}$ over the same range, with the colours indicating the sign of $B_r$. Panels (c) and (g) show $B_r(R_\odot)$ (saturated at $25\,{\rm G}$), while panels (d) and (h) show potential-field source-surface extrapolations taken from this $B_r$ distribution. The field-lines are coloured by $|{\bmath B}|$. \label{fig:cycleinterior}}
\end{figure*}

At a conceptual level, the solar cycle is understood as a process that alternates between poloidal and toroidal field phases.  These phases are connected by sources of poloidal or toroidal magnetic field that help transform one type of magnetic field into the other (while injecting energy into the system).  Current understanding attributes the generation of toroidal field to the shearing of poloidal field by differential rotation, but there is still uncertainty as to which mechanism (or mechanisms) are playing a critical role in the generation of poloidal field out of toroidal field \citep[see review by][for a comprehensive list of the different poloidal sources proposed so far]{Charbonneau2010}.

So far, we have demonstrated that our eruption model is in qualitative agreement with more detailed simulations of individual emergent flux tubes, and that it is capable of producing surface BMRs in agreement with observations.  Now we use this framework to make a global simulation of the solar magnetic field involving hundreds of flux-tube eruptions across an entire solar cycle. This places us in a privileged position to better understand the emergence and decay of BMRs as the main mechanism for creating poloidal field out of toroidal field, i.e. the BL mechanism.

For this purpose,  we have used the locations and times of observed BMRs (from 1996 to 2008) in the NSO/KP dataset from Section \ref{sec:params} to initiate flux-rope eruptions from an underlying sheath of toroidal field at the bottom of the convection zone.  This approach has the added advantage of laying the foundations for the assimilation of BMR data into our model, which is necessary for the study of observed cycle properties and the practical application of models to solar cycle prediction.

\subsection{Setup}

We use an initial condition comprising of (i) a toroidal field belt of strength $B_0=250\,{\rm G}$  (as in Eq.~\ref{eqn:belt}), and (ii) a poloidal field generated by a confined dipole of the form
\begin{equation}
{\bmath B}=\nabla\times(A_\phi\hat{\bmath e}_\phi), \qquad A_\phi=B_{\rm d}\frac{\sin\theta}{r^3}\left(\frac{r-0.7R_\odot}{R_\odot-0.7R_\odot}\right),
\end{equation}
where $A_\phi$ is set to zero for $r<0.7R_\odot$ \citep[c.f.][]{Jouve2008}. The strength of the dipole is set to $B_{\rm d}=-0.008B_0$. The grid resolution in this simulation was set to $\Delta_\phi=2\pi/384$, $\Delta_r=0.45R_\odot/48$, and the same boundary conditions were used as in Section \ref{sec:bcs}.

During the evolution (from 1996 to 2008), new flux-rope emergences were initiated at the times and latitude-longitude locations of observed BMRs (irrespective of the local field strength at the tachocline). The same initial width of $\delta_0=(5\pi/180)(0.7R_\odot)$ was used for all velocity perturbations, chosen because it was the minimum size resolved adequately at our chosen grid resolution (Fig. \ref{fig:sepflux}). Accordingly, only those observed BMRs with flux greater than $10^{22}\,{\rm Mx}$ were included, resulting in 168 BMRs in the Northern hemisphere and 186 in the Southern hemisphere. The same value $\omega_0=0.08\times 10^{-5}\,{\rm s}^{-1}$ was used for all flux tubes (Section \ref{sec:params}), rather than attempting to reproduce the observed scatter of tilt angles about Joy's Law.

\begin{figure*}
\includegraphics[width=140mm]{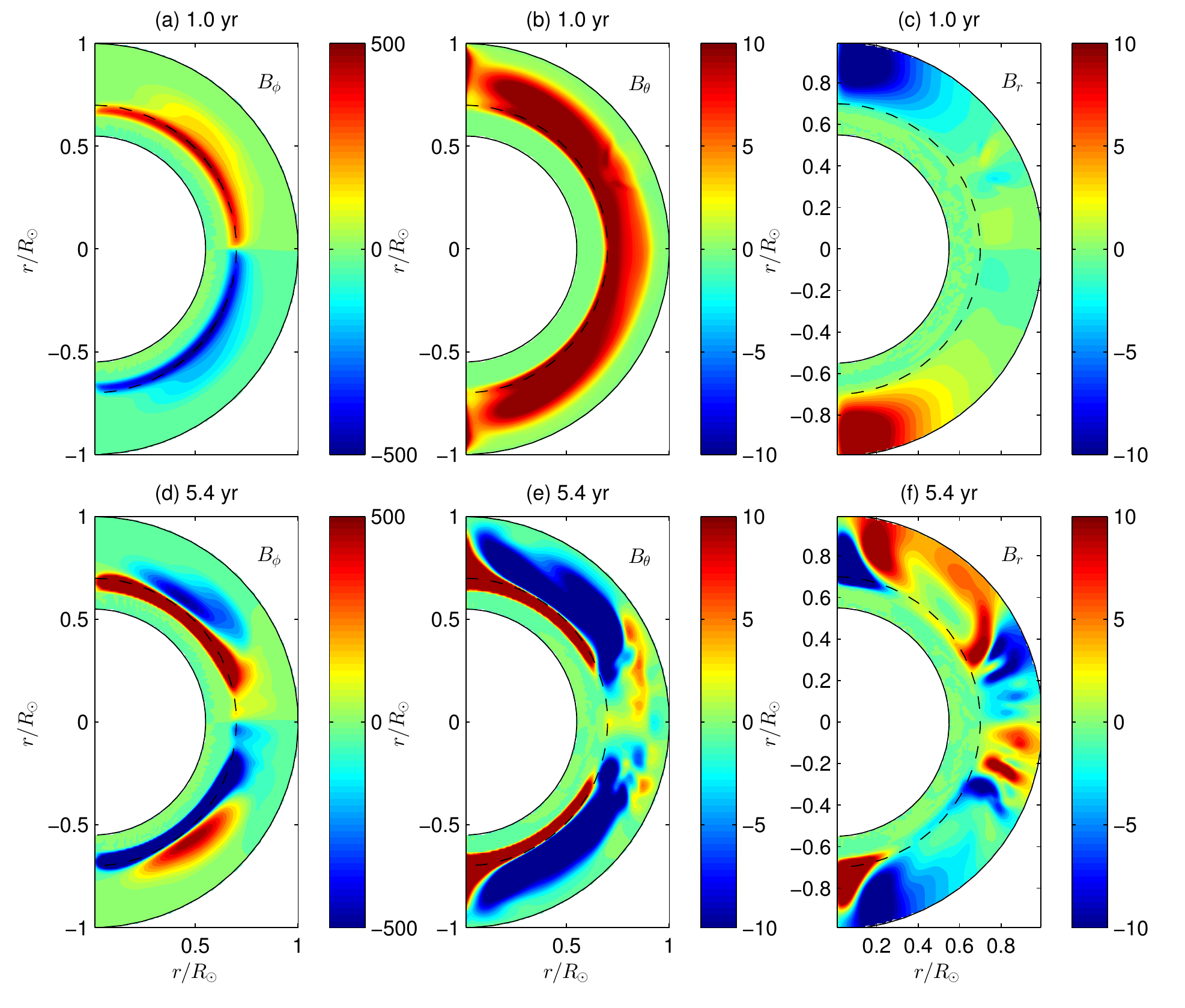}
\caption{Longitude-averaged components of ${\bmath B}$ at the same two times as in Fig. \ref{fig:cycleinterior}. All colour axes are in Gauss and are saturated to show the distribution of weaker magnetic field. Actual maximum values (at 5.4 yr) are $\sim3000\,{\rm G}$ for $B_\phi$, $\sim1000\,{\rm G}$ for $B_\theta$, and $\sim 50\,{\rm G}$ for $B_r$. \label{fig:avmerid}}
\end{figure*}

\subsection{Global magnetic configuration}

Figure \ref{fig:cycleinterior} illustrates the resulting magnetic field configuration at two particular times: early in the simulation when few flux tubes have emerged (top row), and then at the peak of emergence activity (bottom row). Early in the simulation, Fig. \ref{fig:cycleinterior}-a shows that the toroidal field $B_\phi$ reflects the initial conditions, with oppositely-signed belts at the base of the convection zone in each hemisphere. However, there is a clear depletion in $B_\phi$ which corresponds to a recently-emerged flux-tube. The poloidal components of this flux-tube are visible in Fig. \ref{fig:cycleinterior}-b, and as a BMR at the surface in Fig. \ref{fig:cycleinterior}-c.  Looking at the longitudinally averaged components of the magnetic field (Fig. \ref{fig:avmerid}-a to c) clearly shows the combination of the dipolar field with the toroidal sheath, and how the first eruption is already starting to give added structure to the magnetic field inside the convection zone.

By comparison, at the maximum phase of emergence activity, the toroidal field shows the coexistence of multiple belts (Fig. \ref{fig:cycleinterior}-e), although there is a generally preferred polarity at active latitudes. These coexisting belts are also evident in the longitudinally averaged toroidal magnetic field (Fig. \ref{fig:avmerid}-d), which shows the next cycle's toroidal belt being formed at high latitudes.  The many flux-tube eruptions that have taken place have interacted to create wide regions of poloidal field (Fig. \ref{fig:cycleinterior}-f). While recently-emerged active regions are localised, older regions are spread out by the background flow. This spreading is reflected also at the photosphere \ref{fig:cycleinterior}-g).  From an axisymmetric point of view (Fig. \ref{fig:avmerid}-e \& f), the consequence is the formation of a large-scale poloidal field of the opposite polarity that gradually displaces the old polarity and will set the stage for the next cycle.

By assuming a current-free corona, we can calculate a first-order approximation to the coronal magnetic field that would result from a given photospheric distribution of $B_r$. Figures \ref{fig:cycleinterior}-d and \ref{fig:cycleinterior}-h show such potential-field extrapolations, using a source-surface at $r=2.5R_\odot$ where ${\bmath B}$ is forced to be radial \citep{Altschuler1969}. The additional magnetic activity leads to a clear difference in coronal magnetic structure between the two epochs; consisting of equatorial streamers embedded in a predominantly bipolar field during minimum, and a more complex array of streamers and closed field lines during maximum.

\subsection{Evolution of toroidal and poloidal magnetic fields}

\begin{figure*}
\includegraphics[width=168mm]{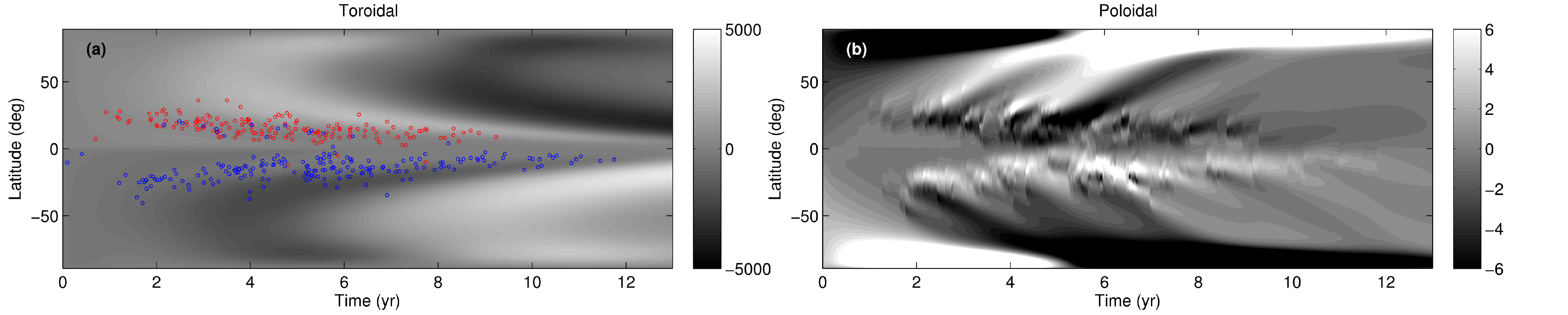}
\includegraphics[width=168mm]{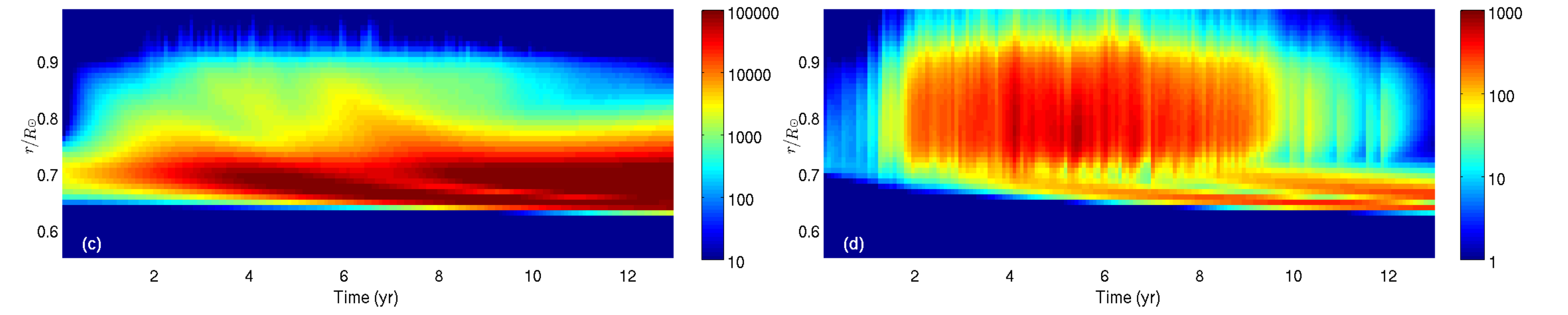}
\caption{Toroidal and poloidal fields in the full solar cycle simulation. Panel (a) shows longitude-averaged $B_\phi$ at $r=0.7R_\odot$, and panel (b) shows longitude-averaged $B_r$ at $r=R_\odot$. Panels (c) and (d) show the radial distributions of toroidal and poloidal magnetic energy density ($B_\phi^2/(8\pi)$ and $(B_r^2 + B_\theta^2)/(8\pi)$ respectively), averaged over $\theta$ and $\phi$, as a function of time. In panel (a), the symbols show the locations of flux-rope eruptions, with the colour denoting the sign of $B_\phi$ at $(\bar{r}(0),\bar{\theta}(0),\bar{\phi}(0))$ (red positive, blue negative). Panels (c) and (d) use a logarithmic colour scale. Units are Gauss in (a), (b), and ${\rm ergs}\,{\rm cm}^{-3}$ in (c), (d).\label{fig:bflyen}}
\end{figure*}

Figure \ref{fig:bflyen} shows the evolution of the toroidal and poloidal magnetic field components over the full simulation. Firstly, Fig. \ref{fig:bflyen}-a shows a longitude-averaged ``butterfly diagram'' of $B_\phi$ at the base of the convection zone, while Fig. \ref{fig:bflyen}-b shows a similar plot of $B_r$ at the surface. The symbols in Fig. \ref{fig:bflyen}-a show the locations of observed BMRs that were used to initiate flux-tube eruptions. Our simulations show that it is possible to achieve good agreement between the time and latitude distribution of observed eruptions and the evolution of the toroidal field at the bottom of the convection zone. The colour of the symbols shows the sign of $B_\phi$ at the initial centre of the velocity perturbation. This generally follows the Hale polarity law; the occasional exceptions are repeat eruptions in locations where the toroidal field has already been removed by a previous eruption, and do not add significant flux to the photosphere.

To better illustrate the interplay between the toroidal and poloidal magnetic fields, we look at the magnetic energy density separated into toroidal and poloidal parts
\begin{equation}
E_{\rm tor} =  \frac{B_\phi^2}{8\pi}, \qquad E_{\rm pol} = \frac{B_r^2 + B_\theta^2}{8\pi}.
\end{equation}
Figures \ref{fig:bflyen}-c and \ref{fig:bflyen}-d show the radial distribution of $E_{\rm tor}$ and $E_{\rm pol}$ (averaged in longitude and latitude) as a function of time, computed from snapshots of ${\bmath B}$ taken at 28-day intervals. Due to the high concentration of toroidal energy at the base of the convection zone, a logarithmic colour-scale is used.  It is evident that throughout the simulation the vast majority of the energy is in the toroidal component; even during the height of surface activity less than $2.5\%$ of the magnetic energy is in the poloidal components of ${\bmath B}$.

Two distinct belts of $E_{\rm tor}$ are apparent at the base of the convection zone in Fig. \ref{fig:bflyen}-c. The first is transported beneath the tachocline allowing the second to form above it from year 8. These are generated by toroidal field belts of \emph{opposite sign}, as is apparent in the toroidal-field butterfly diagram (Fig. \ref{fig:bflyen}-a). The early formation of the second belt may also be seen in Figure \ref{fig:avmerid}-d, which shows the longitude-averaged $B_\phi$ component after 5.4 yr. This second belt has opposite sign to the initial toroidal field, and is generated by differential rotation of new poloidal field created by the flux-tube emergences.  Indeed, the re-generation of toroidal field by decaying BMRs is visible in Fig. \ref{fig:bflyen}-c, particularly during the rising phase. There is an evident creation of $E_{\rm tor}$ within the convection zone, which is subsequently transported downward to the tachocline (by diffusion and pumping).

In contrast to $E_{\rm tor}$, which has a rather smooth distribution, $E_{\rm pol}$ tends to appear in ``bursts'' as flux tubes emerge, before decaying.  The time-scale of this decay is roughly one year, as can be seen in Fig. \ref{fig:bflyen}-d following the final, lone, flux tube emergence in year 11. This decay is partly due to diffusion in the convection zone. However, the decay takes place more rapidly than the diffusion time-scale, which for a structure of size $0.1R_\odot$ at rate $\eta_0$ is about 10 yr. The additional decay is caused by differential rotation, which shears poloidal field to make it toroidal again (thus putting energy back into the toroidal component).

In addition to the main concentration of $E_{\rm pol}$ within the convection zone, Fig. \ref{fig:bflyen}-d shows two downward-propagating bands of $E_{\rm pol}$ that coincide with the two toroidal field belts at the base of the convection zone. In fact, these are two oppositely-signed bands of $B_\theta$ located at high latitudes, visible in Fig. \ref{fig:avmerid}-e when the second band has started to form.  These bands of $B_\theta$ are created at the high-latitude tachocline from earlier build up of strong $B_r$ \emph{of the opposite sign}.  This $B_r$ is located near the pole at all depths in the convection zone (manifested as the polar field at the surface).  The bottom ends of these radial field lines are advected equatorward with respect to the surface (due to meridional flow and turbulent pumping), thus creating a $B_\theta$ component of opposite sign in the tachocline. This latitudinal field is in turn sheared by differential rotation to generate a toroidal component $B_\phi$.

\subsection{Toroidal and poloidal source separation}

\begin{figure*}
\centering
\begin{tabular}{cc}
\includegraphics[width=168mm]{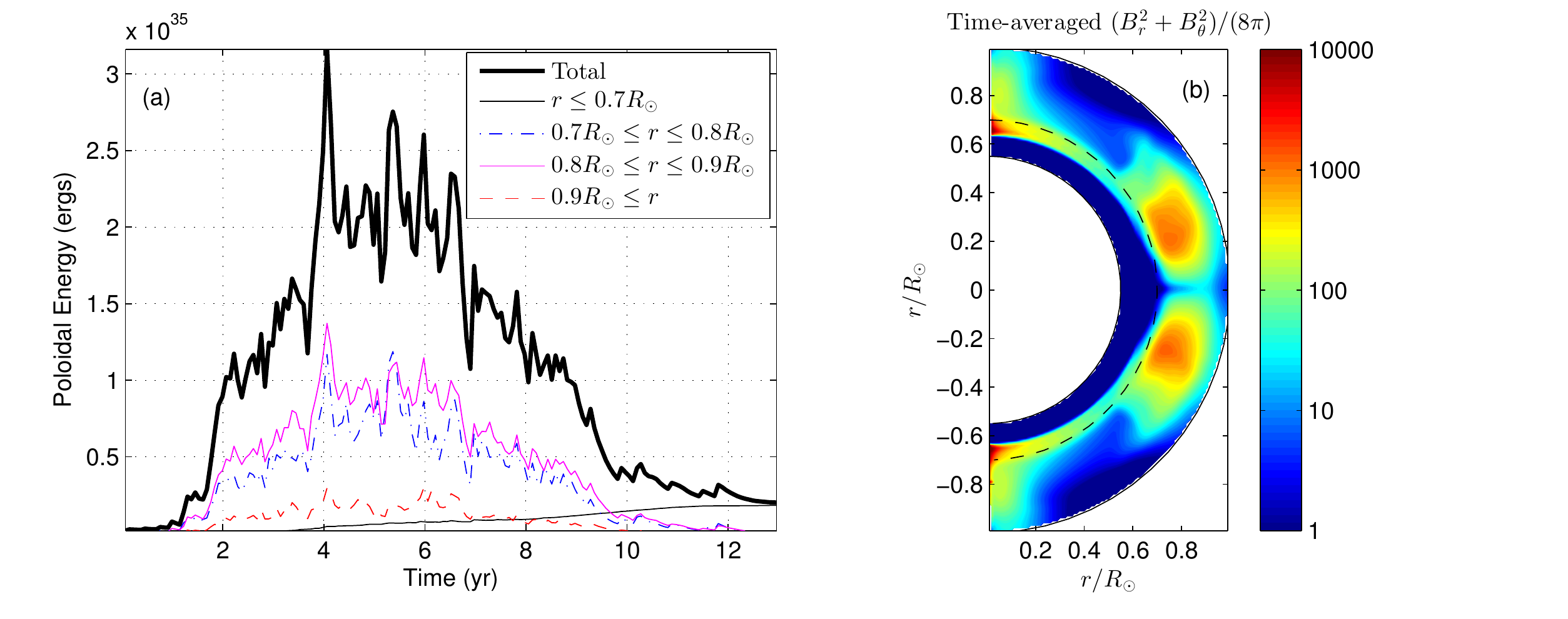}
\end{tabular}
\caption{Evolution of poloidal magnetic energy $E_{\rm pol}$ at different depths in the full solar cycle simulation. In (a), the thick curve shows the total $E_{\rm pol}$ integrated over the full computational domain ($0.55R_\odot\leq r\leq R_\odot$), while the other curves show integrals over successive radial shells (see legend). Panel (b) shows a longitude and time average of the poloidal magnetic energy density ( in ${\rm ergs}\,{\rm cm}^{-3}$, logarithmic scale). \label{fig:enpol}}
\end{figure*}

\begin{figure*}
\centering
\includegraphics[width=168mm]{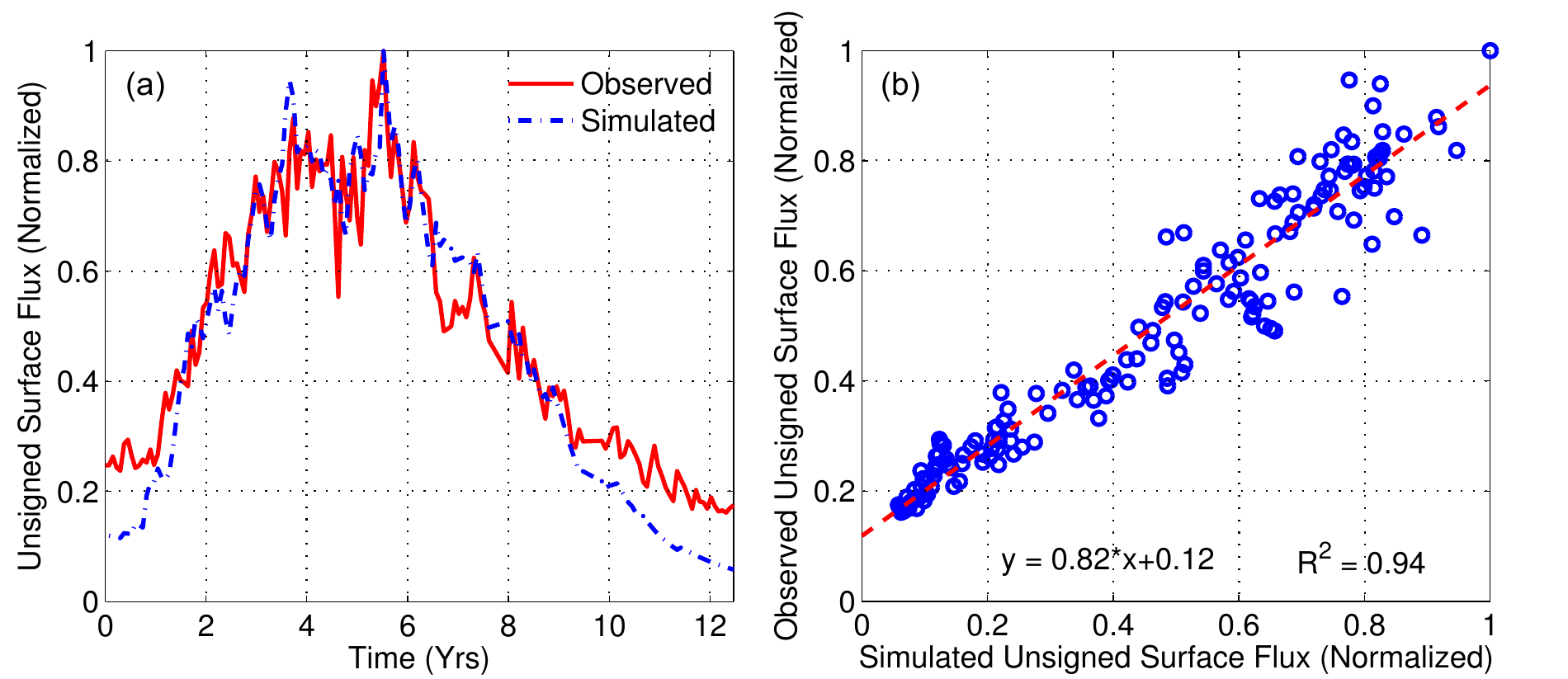}
\caption{Comparison between simulated and observed (NSO/KP) surface magnetic flux, showing (a) normalized fluxes against time, and (b) scatter plot of observed against simulated fluxes. Both fluxes are normalized to unity.\label{fig:simobs}}
\end{figure*}

As can be seen in Figs. \ref{fig:bflyen}-c and \ref{fig:bflyen}-d, the evolution of the toroidal $E_{\rm tor}$ and poloidal $E_{\rm pol}$ energies have rather different radial distributions. We find that whereas the toroidal energy $E_{\rm tor}$ is concentrated in belts at the base of the convection zone, $E_{\rm pol}$ is the strongest in the region containing the legs of the rising tubes (around $r=0.75R_\odot$ to $r=0.9R_\odot$), and weaker both at the bottom of the convection zone and at the surface.  Weak poloidal fields near the bottom of the convection zone are a consequence of the gradual development of tilt in the rising flux tube as it is transported through the convection zone.  Weak poloidal fields near the photosphere are a consequence of the flux-tube expansion and the enhancement of diffusivity in the top-most part of the convection zone (a common feature of modern mean-field dynamo models).  The result is a BL poloidal source that is localized in the midst of the convection zone, rather than at the surface as it is commonly assumed.

Further evidence of the localization of the BL source of poloidal field can be seen in the left panel of Fig. \ref{fig:enpol}, showing the time evolution of the integrated poloidal energy in the lower, middle and top thirds of the convection zone.  During most of the cycle, poloidal energy is largely located in the lower two thirds of the convection zone; with roughly 35\% and 45\% of the total in the lower and middle thirds of the convection zone, and only about 7\% in the top third.  A time and latitudinal average of poloidal energy during the entire cycle (see Fig. \ref{fig:enpol}-b), shows how this concentration of poloidal energy would look in an axisymmetric simulation. Note that the localization of the BL source does not match the profile assumed conventionally \citep{Mason2002f,Dikpati2004b,Guerrero2008,Yeates2008,Mann2009}.

\subsection{Relationship between surface and internal magnetism}

From a practical point of view, one of the most appealing features of the BL mechanism, as the main source of poloidal field, is the strong link it establishes between between observable phenomena at the photosphere and the global evolution of the solar magnetic field.  This gives the surface of the Sun a crucial role in the progression of the solar cycle, instead of simply responding to internal magnetic processes that are currently unobservable.  If this is true, then our data-driven simulation can be used to gain insight into the relationship between surface and interior magnetic field.

A comparison between the observed and simulated total unsigned flux (see Fig. \ref{fig:simobs}-a), shows that total unsigned flux in our simulation is in good agreement with observations (down to the presence of specific features, if not their relative strength).  Note that since arbitrary changes in the strength of the initial toroidal field would allow the simulation to match the amplitude of the observed unsigned flux at the photosphere, we are showing both fluxes normalized.  The main difference between the two is a higher base level in observations compared to simulations.  This difference is to be expected, given that we only use data from BMRs with fluxes above $10^{22}\,{\rm Mx}$.  Figure \ref{fig:simobs}-b shows a scatter plot of observed vs.\ simulated total unsigned flux where this offset is evident.  A linear fit of the relationship between the result of the simulations and observations suggests that the top 21\% of all BMRs (by flux) in the \citet{Yeates2013b} dataset can be used to account for 87\% of the cyclic modulation of total unsigned flux, but more BMRs need to be included to account for the remaining 13\%.

\begin{figure}
\centering
\includegraphics[width=84mm]{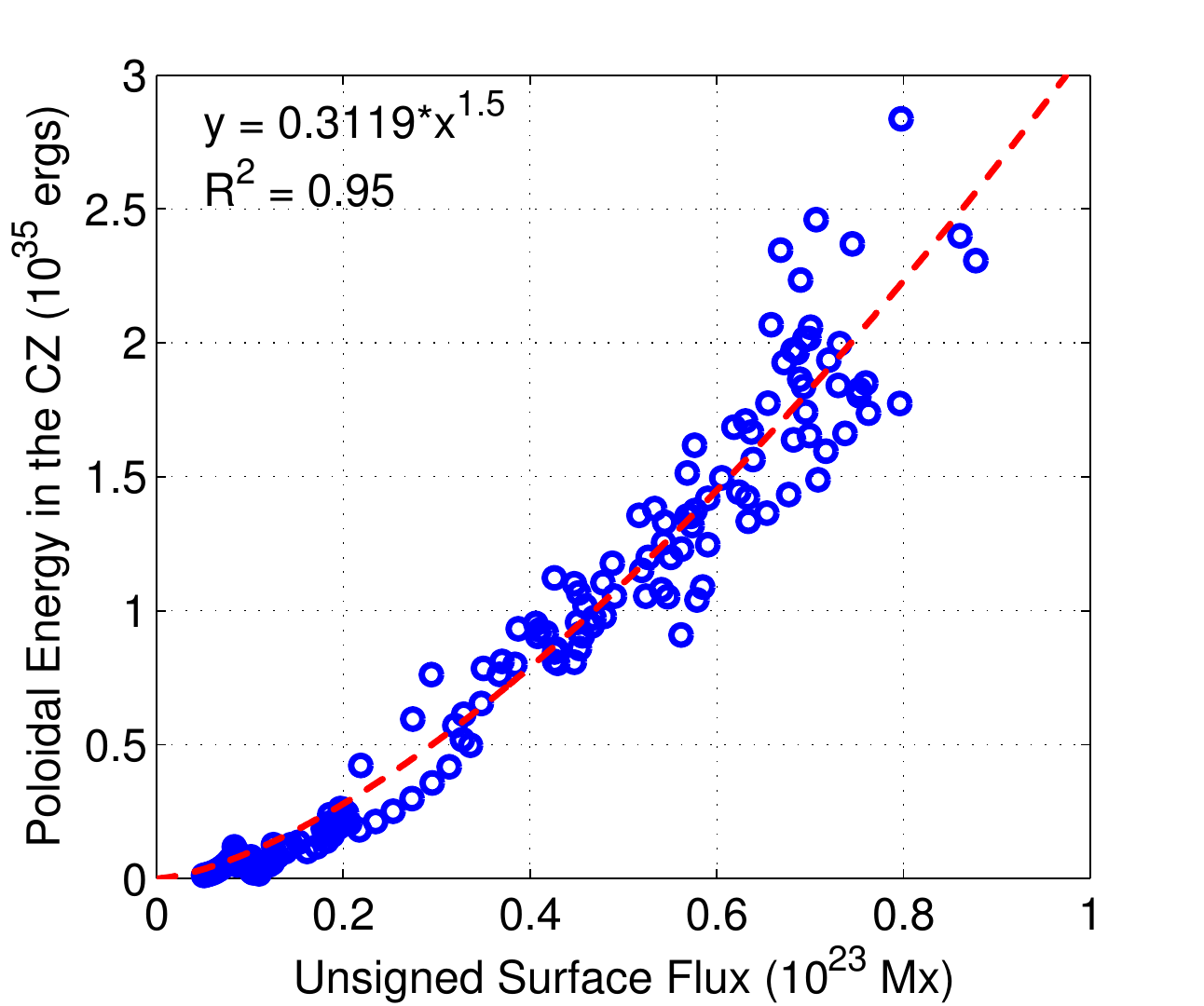}
\caption{Scatter plot of poloidal energy $E_{\rm pol}$ inside the convection zone against total unsigned surface flux, for the full solar cycle simulation. \label{fig:enflux}}
\end{figure}

After demonstrating that our data-driven simulation is in agreement with observations, we finally turn to the relationship between surface and internal magnetism.  The question is whether the magnetic fields observed at the surface are a good indicator of what is happening underneath.  Figure \ref{fig:enflux} shows a scatter plot of total poloidal energy inside the convection zone against total unsigned surface flux. We find that there is a clear monotonic relationship between them, described well by a power law (with a goodness of fit $R^2=0.95$).  We take this as an indicator that, in spite of the relative weakness of surface magnetism, it is a good indicator (in broad terms) of what is going on underneath.

\section{Conclusions} \label{sec:conclusions}

In this Paper, we have introduced a new technique for modelling the emergence of flux-tubes in three-dimensional kinematic dynamo models. We have shown that an appropriate velocity perturbation can reproduce the main observed features of bipolar magnetic regions (BMRs) at the photosphere.  Furthermore, by emerging flux-tubes through this perturbation, our technique avoids the problems associated with artificial flux deposition -- that has previously been used to model the emergence of BMRs -- such as uncertainties in the process of flux removal, the underlying shape of BMRs, and the sudden insertion of flux in the upper convection zone.  However, as we mentioned before, the real objective of this work is not to supersede detailed simulations of single emerging flux-tubes (both using thin flux-tube and anelastic MHD approximations), but rather to incorporate their results into a global simulation where a multitude of emergent flux-tubes during a solar cycle help shape the evolution of the solar magnetic field.

\edit{The results presented in Section \ref{sec:cycle} are the outcome of a data-driven simulation where the time, location, and properties of observed BMRs are used to generate the simulated pattern of eruptions. However, this represents only the first step towards the development of a three-dimensional self-excited dynamo model, where the locations of emerging flux tubes are chosen based on the distribution of magnetic field at the bottom of the convection zone. In future, we intend to extend the full solar cycle simulation presented in Section \ref{sec:cycle} to multiple cycles.  The main objective will be to determine the necessary criteria for establishing a viable set of eruptions leading to a self-propagating cycle. Another important outstanding issue is the question of whether a model incorporating flux emergence of only large BMRs is able to sustain a self-excited cycle. Surface flux transport simulations performed by \citet{Wang1991a} suggest that the largest active regions are the strongest determinants of the evolution of the dipolar field. However, if they are not sufficient to produce a self-excited cycle, it will be necessary to improve the capability of our model to simulate small BMRs and/or include additional sources of poloidal field.}


The more realistic flux-tube emergence process in our model leads to some interesting conclusions. In particular we find that the pinching and reconnection of a flux-tube's legs can occur due to rotational shear. Theoretical considerations suggest that emerging flux-tubes need to disconnect from their roots deep down into the convection zone, in order to explain the observed tilts and surface evolution of BMRs \citep{Fan1994,Longcope2002}. A possible mechanism for disconnection is the debilitating effect of the flux-tube's magnetic field due to pressure and temperature changes in the rising flux-tube in reference to the surrounding media \citep{Schussler2005}.  However, our results suggest that rotational shear can also in principle promote the creation of a reconnection site near the bottom of the convection zone where the disconnection can take place; a process that could be further enhanced by converging flows caused by the eruption of magnetic flux from the stable layer into a superadiabatically stratified convection zone \citep[see][]{Hotta2012}.  It will be interesting if detailed anelastic MHD simulations of the emergence of a buoyant flux tube into a region of strong radial shear can validate this idea; as the results of \cite{Jouve2013} seem to suggest.

Next, our simulation of a full solar cycle with multiple tubes challenges the frequent assumption in many flux-transport dynamo models that the poloidal field source is located at (or very near to) the solar surface. In fact, we find that the poloidal magnetic field is strongest not at the surface but in the middle of the convection zone, among the legs of the rising flux tubes. Our results suggest that the physical separation of sources is not radial, but latitudinal (compare Figs. \ref{fig:avmerid}-d \ref{fig:enpol}-b).  This means that the mechanisms responsible for latitudinal flux transport, together with the process of flux removal and cancellation, are probably the most important for setting the cycle timing. Future work will need to assess the consequences of this shortening of the flux-transport loop.

Additionally, we find good agreement between the simulated and observed surface evolution of the magnetic field. If this agreement can be extended to the rest of the convection zone, then the energy contained in the near-surface magnetic field is much weaker than that in the interior.  However, we find surface magnetism to be well correlated with the evolution of the internal magnetic field, suggesting that surface observations can be used as an indicator of the state of the magnetic field in the interior of the Sun.

Finally, our new technique for the eruption of flux-tubes in a kinematic framework, and our method for BMR data assimilation, lay the foundations for a new generation of kinematic models that will prove instrumental for furthering our understanding of the solar cycle and improving our capabilities for cycle prediction.  However, in contrast to the previous generation of axisymmetric models, the three-dimensional nature of our simulations will allow our dynamo model to couple to three-dimensional simulations of the Sun's coronal magnetic field (as demonstrated in Figs. \ref{fig:cycleinterior}-d and \ref{fig:cycleinterior}-h), paving the way for the simultaneous study of the evolution of the magnetic field in the solar interior as well as its impact on the heliosphere.

\section*{Acknowledgments}

AM-J is grateful to David Kieda for his support and sponsorship at the University of Utah. AM-J is supported by the NASA Living With a Star Jack Eddy Postdoctoral Fellowship Program, administered by the UCAR Visiting Scientist Programs.  Numerical simulations used the STFC and SRIF funded UKMHD cluster at the University of St Andrews. Figures \ref{fig:3d}-a and \ref{fig:cycleinterior} used the NCAR VAPOR visualization software ({www.vapor.ucar.edu}). Magnetogram data from NSO/Kitt Peak were produced cooperatively by NSF/NSO, NASA/GSFC,and NOAA/SEL, and SOLIS data are produced cooperatively by NSF/NSO and NASA/LWS.

\bibliographystyle{mn2e}
\bibliography{ary_dynamo}

\appendix

\section{Numerical Methods} \label{sec:num}

To solve Eq. (\ref{eqn:inductionB}) numerically, we write ${\bmath B}=\nabla\times{\bmath A}$ and solve for the fully three-dimensional vector potential ${\bmath A}(r,\theta,\phi,t)$ in spherical coordinates. We choose the gauge of ${\bmath A}$ so that
\begin{equation}
\frac{\partial{\bmath A}}{\partial t}=-\bmath{\mathcal E},
\label{eqn:inductionA}
\end{equation}
where $\bmath{\mathcal E}$ is the mean electromotive force
\begin{equation}
\bmath{\mathcal E} = -{\bmath v}\times{\bmath B} + \eta\nabla\times{\bmath B}.
\label{eqn:emf}
\end{equation}
Equation \eqref{eqn:inductionA} is solved for ${\bmath A}$ using finite-differences in a spherical shell  $R_{\rm min}<r<R_\odot$, $\theta_\textrm{min}<\theta<\pi-\theta_\textrm{min}$, $0\le\phi <2\pi$, where we set the latitudinal boundaries near to the pole, e.g. $\theta_\textrm{min}=0.5\degr$. The pole itself is incorporated through special boundary conditions on this boundary, where we set $B_\phi=0$ and choose $B_r$ to satisfy Stokes' Theorem given the integral of $A_\phi$ around the latitudinal boundary at each radius. To minimise numerical error in the advection terms, we solve in a frame rotating at constant angular velocity $\Omega_C$ (the induction equation remains unchanged).

Rather than solving in $(r,\theta,\phi)$ coordinates, we use stretched coordinates $(x,y,z)$ which are not Cartesian but are defined by
\begin{align}
x &= \phi/\Delta_\phi,\\
y &= -\log\left(\tan\left(\theta/2 \right) \right)/\Delta_\phi,\\
z &= (r-R_\textrm{min})/\Delta_r,
\end{align}
where $\Delta_\phi$ is the azimuthal cell size (in radians) at the equator, and $\Delta_r$ is the radial cell size \citep[cf.,][]{VanBallegooijen2000}. These coordinates have the advantage that the horizontal coordinate scale-factors are equal, $h_x = h_y = \Delta_\phi r \sin\theta$.

\begin{figure}
\includegraphics[width=68mm]{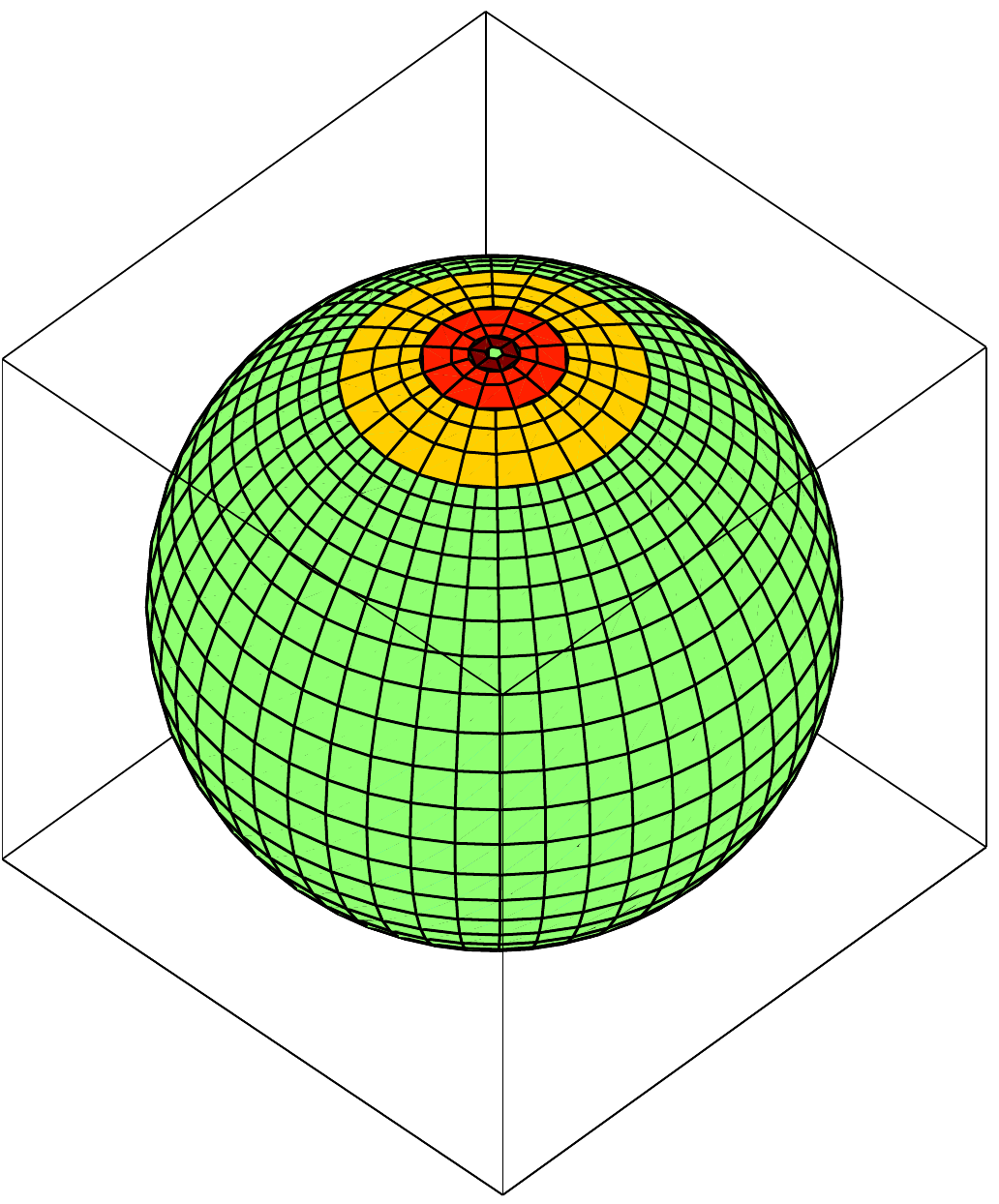}
\caption{The variable grid in $(x,y)$, for $\Delta_\phi=2\pi/48$. Different sub-blocks toward the north pole are indicated by colour shading. \label{fig:grid}}
\end{figure}

Furthermore, to avoid severe time-step restrictions caused by convergence of the grid points near the poles (the horizontal cell area is $\Delta^2_\phi r^2 \sin^2\theta$), we adopt a variable grid spacing in $x$, $y$. The grid is decomposed into individual sub-blocks in the latitudinal direction (Fig. \ref{fig:grid}). Within each sub-block, the cell sizes $dx$, $dy$ are constant. For the sub-block covering the equator, $dx=dy=1$, while $dx$ and $dy$ are doubled in each poleward sub-block. The sub-block boundaries are placed at the lowest latitude possible such that no cell has area $h_xh_y\,dxdy$ greater than the equatorial cell area $\Delta_\phi^2r^2$. All subgrids have $dz=1$. For example, with a longitudinal resolution of $\Delta_\phi=2\pi/192$ at the equator, we obtain 9 sub-blocks and a total of 18\,936 grid cells in $(x,y)$, as opposed to 63\,744 for a uniform single-block grid. Alternative solutions to the grid convergence problem include the use of spectral methods, unstructured grids, or of multiple overset grids such as the Yin-Yang \citep{Kageyama2004} or cubed-sphere \citep{Ronchi1996} grids. Our method has the advantage, since we used the constrained-transport formulation, of readily incorporating magnetic flux conservation at the sub-block boundaries.

Within each sub-block, we employ the constrained transport formalism \citep{evans1988}, whereby the components of ${\bmath B}$ are treated as fluxes through the $x,y,z$ cell faces. These are evolved by $\bmath{\mathcal E}$ on the cell edges, where the components of ${\bmath A}$ are also located. The advection terms ${\bmath v}\times{\bmath B}$ are treated using an upwind average of ${\bmath B}$ to the edges. We use the superbee flux limiter to obtain second-order accuracy away from sharp jumps in ${\bmath B}$ while switching to the first-order donor cell scheme at sharp jumps, thus preventing spurious oscillations. Time-stepping uses the second-order trapezoidal scheme, which requires two evaluations of $\bmath{\mathcal E}$ per time-step. If ${\bmath A}^n$ denotes the value of ${\bmath A}$ at $t=t_n$, then the value of ${\bmath A}^{n+1}$ at $t_{n+1}=t_n+dt$ is obtained in two stages by
\begin{align}
\widetilde{\bmath A} &= {\bmath A}^n + dt\,\bmath{\mathcal E}({\bmath A}^n,t_n),\\
{\bmath A}^{n+1} &= {\bmath A}^n + \frac{dt}{2}\bmath{\mathcal E}({\bmath A}^n,t_n) + \frac{dt}{2}\bmath{\mathcal E}(\widetilde{\bmath A},t_{n+1}).
\end{align}

The variable grid requires two modifications at sub-block boundaries. Firstly, ghost-cell values of $B_x$ and $B_z$ need to be transferred between neighbouring sub-blocks. This is analogous to the inter-level communications in AMR (adaptive mesh refinement) codes, except that our grid is fixed in time. The simpler process is ``restriction'', for which we use an area-weighted average of finer-grid values to give the ghost-cell values on the coarser grid \citep{Balsara2001h}. To obtain ghost-cell values on the finer grid, we need to interpolate values on the coarser grid, a process called prolongation \citep{Balsara2001h,Toth2002}. We follow a method similar to \citet{VanDerHolst2007} based on Taylor expansion. The other modification required is a correction to $\bmath{\mathcal E}$ on the boundaries of coarser grids  before updating ${\bmath A}$, to ensure that it matches the values computed by finer grids on the same boundary. This is analogous to the ``flux correction'' of \citet{Berger1989}, and done in a weighted sense.

The code is parallelized with OpenMPI and the simulation described in Section \ref{sec:cycle}, with $\Delta_\phi=2\pi/384$, $\Delta_r=0.45R_\odot/48$ took 6.25 hours on a modest 48 cores. In particular, it will be practical to run simulations of multiple solar cycles at this resolution.

\bsp

\label{lastpage}

\end{document}